\documentclass[a4paper,11pt]{article}
\pdfoutput=1 \usepackage{jheppub_2} 
\usepackage{booktabs}
\usepackage[T1]{fontenc} 
\usepackage{color,subfigure}
\usepackage{graphicx, multirow,soul,url,amsmath,amsfonts,amssymb,mathrsfs,amsfonts}
\usepackage[all]{hypcap}
\usepackage{url}
\usepackage{bbm}
\usepackage{cancel}
\usepackage{subfigure}
\usepackage{slashed}
\usepackage[dvipsnames]{xcolor}
\usepackage{multicol, blindtext}
\usepackage[section]{placeins}
\usepackage[version=3]{mhchem}
\usepackage{caption}
\usepackage{lipsum}
\usepackage{hyperref}
\usepackage{float}
\usepackage{mattens}
\usepackage{mathtools}
\usepackage{footnote}
\usepackage{amsmath,amssymb,amsthm,amsxtra,overpic,bbm,bm,epsfig}
\usepackage{color,subfigure}
\usepackage[splitrule]{footmisc}
\usepackage{bbold}
\usepackage{amsmath}
\usepackage{amsbsy}
\usepackage{lipsum}
\usepackage[export]{adjustbox}
\usepackage{units}
\newcommand{\overbar}[1]{\mkern 1.1mu\overline{\mkern-1.1mu#1\mkern-1.1mu}\mkern 1.1mu}
\newcommand{\equaref}[1]{Eq.~(\ref{#1})}

\newcommand{\figref}[1]{Fig.~\ref{#1}}

\newcommand{\secref}[1]{Section~\ref{#1}}

\newcommand{\tabref}[1]{Table~\ref{#1}}

\newcommand{\bq}{\begin{eqnarray}}
\newcommand{\nq}{\end{eqnarray}}

\title{\bf Leptogenesis in the Neutrino Option}
\author[a]{I. Brivio,}
\affiliation[a]{Institut f{\"u}r Theoretische Physik, Universit{\"a}t Heidelberg, Philosophenweg 16, 69120 Heidelberg,
Germany.}
\author[b]{K. Moffat,}
\author[b]{S. Pascoli,}
\affiliation[b]{Institute for Particle Physics Phenomenology, Department of
Physics, Durham University, South Road, Durham DH1 3LE, United Kingdom.}
\author[c,d]{S.T. Petcov}
\affiliation[c]{SISSA/INFN, Via Bonomea 265, I-34136 Trieste, Italy.}
\affiliation[d]{Kavli IPMU (WPI), University of Tokyo, 5-1-5 Kashiwanoha, 277-8583 Kashiwa, Japan.}	
\author[e]{and J. Turner}
\affiliation[e]{Theoretical Physics Department, Fermi National Accelerator Laboratory, P.O. Box 500, Batavia, IL 60510, USA.}

\emailAdd{brivio@thphys.uni-heidelberg.de}
\emailAdd{kristian.p.moffat@durham.ac.uk}
\emailAdd{silvia.pascoli@durham.ac.uk}
\emailAdd{petcov@sissa.it}
\emailAdd{jturner@fnal.gov}

\abstract{We examine the compatibility between the Neutrino Option, 
in which the electroweak scale is generated by PeV mass type I seesaw 
Majorana neutrinos, and leptogenesis. We find the Neutrino Option is consistent 
with resonant leptogenesis. Working within the minimal 
seesaw scenario with two heavy Majorana neutrinos $N_{1,2}$, which
form a pseudo-Dirac pair,
we explore the viable parameter space.
We find that the Neutrino Option and successful leptogenesis are compatible 
in the cases of a neutrino mass spectrum with normal (inverted) ordering
for $1.2 \times 10^6 < M \text{ (GeV)} < 8.8 \times 10^6$ 
($2.4 \times 10^6 < M \text{ (GeV)} < 7.4 \times 10^6$), 
with $M = (M_1 + M_2)/2$ and $M_{1,2}$ the masses of $N_{1,2}$.
Successful leptogenesis requires 
that $\Delta M/M \equiv (M_2 - M_1)/M \sim 10^{-8}$. 
We  further show that leptogenesis can produce the baryon asymmetry of the Universe within 
the Neutrino Option scenario  when the requisite CP violation 
in leptogenesis is provided exclusively by 
the Dirac or Majorana low energy CP violation 
phases of the PMNS matrix.
}

\preprint{
\begin{flushleft}{
FERMILAB-PUB-19-248-T\\
IPMU19-0084\\
SISSA 15/2019/FISI\\
IPPP/19/47}
\end{flushleft}
}


\begin{document}

\thispagestyle{empty}
\def\thefootnote{\fnsymbol{footnote}}
\setcounter{footnote}{1}

\setcounter{page}{0}
\maketitle
\vspace{-1cm}
\flushbottom

\def\thefootnote{\arabic{footnote}}
\setcounter{footnote}{0}

\newpage
\section{Introduction}\label{sec:introduction}
Neutrino oscillation experiments have provided overwhelming evidence 
for very small but non-zero neutrino masses --- a fact that is not 
explained in the Standard Model (SM). 
A minimal extension of the SM which addresses this issue is 
the type~I seesaw framework where a number of heavy Majorana 
neutrinos are added to the SM particle spectrum
~\cite{Minkowski:1977sc,Yanagida:1979as,GellMann:1980vs,Mohapatra:1979ia}. 
The light neutrinos have a mass scale inversely proportional to that 
of the heavy Majorana neutrinos and so their extreme smallness may be 
naturally explained. In particular, if the type~I seesaw mechanism is 
to generate at least 
three non-degenerate
light neutrino masses, as is required by the data 
from oscillation experiments, then there must be at least two 
heavy Majorana neutrinos.

The decays of the heavy Majorana neutrinos in the early Universe 
may produce the observed matter-antimatter asymmetry in leptogenesis  
processes~\cite{Fukugita:1986hr} 
(see also, e.g., \cite{Buchmuller:1996pa,Buchmuller:2004nz,Buchmuller:2005eh,Pilaftsis:1997jf}).
 Lepton-number- and C- and CP-violating decays of the 
heavy Majorana neutrinos, which generate a lepton asymmetry,
may occur out-of-equilibrium 
thereby satisfying Sakharov's conditions
 for a dynamical generation of the (leptonic) matter-antimatter 
asymmetry of the Universe \cite{Sakharov:1967dj}. This lepton asymmetry is subsequently partially processed into a 
baryon asymmetry by $B+L$-violating sphaleron processes.

However, there exists a tension between leptogenesis within the type I seesaw mechanism
and the naturalness of the Higgs potential. This is because 
radiative corrections to the Higgs potential increase 
monotonically with the mass scale of the heavy Majorana neutrinos. 
In a natural scenario , where the corrections to the Higgs mass 
do not exceed $1$ TeV, 
the heavy Majorana neutrino mass scale must satisfy 
$M < 3 \times 10^7$ GeV~\cite{Vissani:1997ys,Clarke:2015gwa}, 
which is considerably lower than 
the Davidson-Ibarra bound for successful leptogenesis $M \gtrsim 10^9$ GeV~\footnote{We recall that the Davidson-Ibarra bound is valid for 
hierarchical heavy Majorana neutrino mass spectrum 
and in the case of absence of flavour effects in leptogenesis.
}
~\cite{Davidson:2002qv,Buchmuller:2002rq,Ellis:2002xg}.

A different perspective on this problem is brought by the so-called Neutrino Option scenario
~\cite{Brivio:2017dfq,Brivio:2018rzm} which is based on
the idea that the Higgs potential is generated by the radiative 
corrections of the heavy Majorana neutrinos,
 starting from an approximately conformal scalar potential at the seesaw scale. In this framework the heavy Majorana neutrino masses are the only dimensionful parameters of the theory and they control both the breaking of the conformal symmetry and that of lepton number.

In this work we shall investigate the possibility of successful 
leptogenesis within the Neutrino Option framework and focus on the 
minimal scenario where there are only two heavy Majorana neutrinos 
providing both the Higgs mass and the baryon asymmetry. We show that in order for leptogenesis to be successful within the Neutrino Option approach to electroweak symmetry breaking,
it is necessary for the two heavy neutrinos to be close in mass (forming a pseudo-Dirac 
pair \cite{Wolfenstein:1981kw,Petcov:1982ya}) and their masses to be 
in the range $M \sim 10^{6}- 10^{7}$ GeV. 
From these considerations, we derive an upper and lower bound on this mass scale. 
The upper bound coincides with 
that found for the Neutrino Option alone~\cite{Brivio:2018rzm}, 
while the lower bound comes 
from the additional requirement of viable leptogenesis. 
We explore the naturalness in the neutrino 
sector in terms of the presence of an approximately conserved lepton charge.
We investigate also the production of the baryon asymmetry 
when the requisite CP violation 
in leptogenesis is provided exclusively by 
the low energy phases of the PMNS matrix.

%
%
\section{The Neutrino Option and Leptogenesis}
\label{sec:framework}

In this section we review neutrino masses in the type I seesaw (\secref{sec:numassandmix}). We then summarise the essential features of the Neutrino Option in \secref{subsec:TheNO} before reviewing the relevant features of leptogenesis in \secref{subsec:LGframework}. Finally in \secref{subsec:LGatNOscales} we consider which leptogenesis scenarios may be viable in conjunction with the Neutrino Option.

%
%

\subsection{Neutrino masses and mixing}
\label{sec:numassandmix}
%
%

 Neutrino oscillation experiments have shown that neutrinos have small 
but non-zero masses and that they mix. 
The mass and flavour states of neutrinos are not aligned and 
their misalignment is described by the PMNS matrix $U$:
\begin{equation}
\nu_{\alpha L} = \sum_{i=1}^3 U_{\alpha i} \nu_{i L},
\end{equation}
%
where $\nu_{\alpha L}$ is the flavour neutrino field,
$\alpha \in \{e,\mu,\tau\}$, and $\nu_{i}$ is the 
field of the neutrino of mass $m_i$.
We make use of the conventional PDG parametrisation~\cite{PDG2018}:
\begin{equation}
U =\begin{pmatrix}
c_{12}c_{13} & s_{12}c_{13} & s_{13} e^{-i \delta} \\
-s_{12}c_{23}-c_{12}s_{23}s_{13} e^{i \delta} & c_{12} c_{23} - s_{12} s_{23} s_{13} e^{i \delta} & s_{23} c_{13}  \\
s_{12}s_{23}-c_{12}c_{23}s_{13} e^{i \delta} & -c_{12}s_{23}-s_{12}c_{23}s_{13} e^{i \delta} & c_{23}c_{13} 
\end{pmatrix}
\begin{pmatrix}
1 & 0 & 0\\
0&e^{i\frac{\alpha_{21}}{2}} & 0\\
0 & 0 &  e^{i\frac{\alpha_{31}}{2}}
\end{pmatrix},
\end{equation}
%
%
where $c_{ij} \equiv \cos\theta_{ij}$, $s_{ij} \equiv \sin\theta_{ij}$, $\delta$ 
is the Dirac  phase and $\alpha_{21}$, $\alpha_{31}$ are the 
Majorana phases~\cite{Bilenky:1980cx}. 
In addition to the mixing angles and phases, 
neutrino oscillation experiments provide precise measurements of the 
two independent neutrino mass squared differences  $\Delta m^2_{21}$ and 
$\Delta m^2_{31(32)}$. 
The best-fit values 
of $\theta_{12}$, $\theta_{23}$,  $\theta_{13}$, $\delta$, 
$\Delta m^2_{21}$ and $\Delta m^2_{31(32)}$, obtained in one of the most 
recent analyses of the global neutrino oscillation data 
\cite{Esteban:2018azc}, are given in \tabref{tab:bestfitdata}.
The three neutrino masses $m_1$, $m_2$ and $m_3$ may be arranged 
into two possible orderings compatible with the oscillation data: 
normal ordering (NO) for $m_1<m_2<m_3$; 
and inverted ordering (IO) $m_3<m_1<m_2$. 
Although the mass ordering has not yet been determined experimentally, 
there is a mild statistical preference for normal ordering 
in the data~\cite{Esteban:2018azc}. 
\begin{table}[t]
\centering
\begin{tabular}{ c  c  c  c  c  c  c }
 \toprule
$\theta_{13}$ & $\theta_{12}$ & $\theta_{23}$ & $\delta$ & $\Delta m_{21}^2$ & $\Delta m_{3l}^2$\\
$(^{\circ})$ & $(^{\circ})$ & $(^{\circ})$ & $(^{\circ})$ & ($10^{-5} \text{eV}{}^2$) & ($10^{-3} \text{eV}{}^2$) \\
\specialrule{2.5pt}{1pt}{1pt}
$8.61^{+0.12}_{-0.13}$ & $33.82^{+0.78}_{-0.76}$ & $49.7^{+0.9}_{-1.1}$ & $217^{+40}_{-28}$ & $7.39^{+0.21}_{-0.20}$ & $2.525^{+0.033}_{-0.031}$ \\
\specialrule{2.5pt}{1pt}{1pt}
$8.65^{+0.12}_{-0.13}$ & $33.82^{+0.78}_{-0.75}$ & $49.7^{+0.9}_{-1.0}$ & $280^{+25}_{-28}$ & $7.39^{+0.21}_{-0.20}$ & $-2.512^{+0.034}_{-0.031}$ \\
\bottomrule
\end{tabular}\caption{Best fit value and 1$\sigma$ ranges 
 of the neutrino oscillation parameters
from a global 
fit to neutrino data \cite{Esteban:2018azc}. 
 Upper (lower) values are for neutrino mass spectrum with 
normal (inverted) ordering, for which $\Delta m_{3l}^2 \equiv \Delta m_{31}^2$ 
($\Delta m_{3l}^2 \equiv \Delta m_{32}^2$).
}
\label{tab:bestfitdata}
\end{table}
%

A simple means of explaining the smallness of neutrino masses 
is the type I seesaw framework~\cite{Minkowski:1977sc,Yanagida:1979as,GellMann:1980vs,Mohapatra:1979ia}, 
where heavy Majorana neutrinos, $N_i$ ($i \in \{1,2,3,\hdots\}$), are added 
to the SM spectrum.  After  electroweak symmetry is broken, when the Higgs has developed 
a vacuum expectation value (vev)  $v \approx 174 \text{ GeV}$, 
the neutrino mass terms of the Lagrangian are given by
\begin{equation}
\label{Mdefinition}
\begin{aligned}
\mathcal{L}_m & = -\frac{1}{2} \left(\bar{\nu}_L, \bar{N}^c_L \right) 
\left(\begin{array}{cc}
0 & v Y\\
v Y^T & M_R \\
\end{array}\right)
\left(\begin{array}{c}
\nu_R^c\\
N_R\\
\end{array}\right)
+\text{h.c.},
\end{aligned}
\end{equation}
%
where $\nu_L^T \equiv ( \nu_{eL}^T, \nu_{\mu L}^T, \nu_{\tau L}^T )$ 
and $N_R^T \equiv \left( N_{1R}^T, N_{2 R}^T, N_{3 R}^T,\hdots \right)$
 are the flavour neutrino fields and the fields of the heavy Majorana 
mass eigenstates respectively and $Y$ is the matrix of the neutrino Yukawa couplings of
the heavy Majorana neutrinos $N_{i}$ to the leptonic and Higgs doublets. 
In \equaref{Mdefinition}, 
$(\nu_R^c)^T = ((\nu^c_{eR})^T, (\nu^c_{\mu R})^T, (\nu^c_{\tau R})^T))$, 
$(N^c_L)^T = \left( (N^c_{1L})^T, (N^c_{2 L})^T, (N^c_{3 L})^T, \hdots \right)$,
with  $\nu^c_{lR} = C(\bar{\nu}_{lL})^T$, $l \in \{e,\mu,\tau\}$, 
and  $N^c_{j L} = C (\bar{N}_{jR})^T$ for $j \in \{1,2,3,\hdots\}$, where
$C$ denotes the charge conjugation matrix. We work in a basis in which the Majorana mass matrix $M_R$ is diagonal $M_R = \text{diag}(M_1,M_2,\hdots)$.

 The light neutrino mass matrix, $m_{\nu}$, may be 
brought to a semi-positive
diagonal form (denoted by a caret) using the Takagi transformation:
\begin{equation}
\hat{m}_{\nu} = U^{\dagger} m_{\nu} U^*,
\end{equation}
%
such that $\hat{m}_{\nu} = \text{diag}(m_1,m_2,m_3)$, with $m_i$ 
the mass corresponding to $\nu_i$.

By analogy with the method of Casas and Ibarra~\cite{Casas:2001sr}, we apply the seesaw formula for light neutrino masses
\begin{equation}
m_\nu \approx - v^2 Y M_R^{-1} Y^T,
\end{equation}
to parametrise the Yukawa matrix as
\begin{equation}
\label{eq:CIparam}
Y=\frac{1}{v}U\sqrt{\hat{m}_{\nu}}R^T\sqrt{M_R},
\end{equation}
%
where $R$ is a $3 \times 3$ complex orthogonal matrix. Here we drop a phase factor of $i$, which would cancel from all our results. We note that this result, which is correct at tree-level, will suffice in this work as radiative corrections to the light neutrino masses will be negligible.

In what follows we consider the case in which only two heavy 
Majorana neutrinos are present, or equivalently, the case when the third 
heavy Majorana neutrino, $N_3$, decouples.
This is the minimal scenario compatible with the oscillation data where 
one of the three light neutrinos is massless. For concreteness, we also assume $M_1\leq M_2$. 
In the considered  scenario,  the sum of 
neutrino masses is $\sum_{i=1}^{3}m_{\nu_{i}} \cong 0.058\,(0.010)\,\text{eV}$ 
for NO (IO) neutrino mass spectrum,
which is well within the cosmological upper limit 
reported by the Planck collaboration, 
$\sum_i m_i < 0.120 - 0.160$ eV at 95\% C.L.~\cite{Aghanim:2018eyx}.

The assumption of $N_3$ decoupling implies that we may take $m_1 = 0$ for the NO spectrum (at tree-level). 
Correspondingly we have for the $R-$matrix:
\begin{equation}
R = 
\begin{pmatrix}
0 & \cos \theta&  \sin \theta \\
0 & - \sin \theta & \cos\theta \\
1& 0 & 0
\end{pmatrix},
\label{eq:R2NNO} 
\end{equation}
%
where  $\theta=x + iy$. This in turn leads to
\begin{equation}
U^\dagger Y = 
\frac{1}{v}
\left(
\begin{array}{cc}
 0 & 0 \\
\sqrt{M_1} \sqrt{m_2} \cos \left( x + i y \right) & - \sqrt{M_2} \sqrt{m_2} \sin \left( x + i y \right)\\
\sqrt{M_1} \sqrt{m_3} \sin \left( x + i y \right) & \sqrt{M_2} \sqrt{m_3} \cos \left( x + i y \right) \\
\end{array}
\right).
\label{eq:CINO}
\end{equation}
%
Similarly, for IO spectrum we have $m_3 = 0$ and 
subsequently
\begin{equation}
R = 
\begin{pmatrix}
 \cos \theta&  \sin \theta & 0\\
- \sin \theta & \cos\theta & 0\\
0 & 0 & 1
\end{pmatrix}\,.
\label{eq:R2NIO} 
\end{equation}
%
In this case
\begin{equation}
U^\dagger Y = 
\frac{1}{v}
\left(
\begin{array}{cc}
\sqrt{M_1} \sqrt{m_1} \cos \left( x + i y \right) & - \sqrt{M_2} \sqrt{ m_1} \sin \left( x + i y \right) \\
\sqrt{M_1} \sqrt{m_2} \sin \left( x + i y \right) & \sqrt{M_2} \sqrt{m_2} \cos \left( x + i y \right) \\
 0 & 0 \\
\end{array}
\right).
\label{eq:CIIO}
\end{equation}
%
In the literature a phase factor $\xi = \pm 1$ is sometimes included 
in the definition of $R$ to allow 
for both the cases $\det(R)=\pm 1$. We have chosen instead to extend the range of the Majorana phases 
$\alpha_{21(31)}$ to $[0,4\pi]$
such that the same full set of possible Yukawa matrices has 
been considered \cite{Molinaro:2008rg}.

%
\subsection{The Neutrino Option}
\label{subsec:TheNO}

\begin{figure}[t]
\centering
\includegraphics[width=0.8\textwidth]{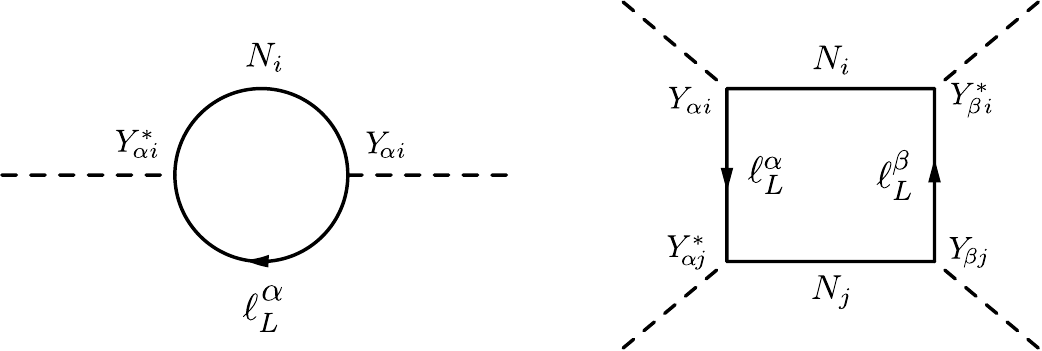}
\caption{The dominant one-loop contribution generating the Higgs potential 
in the type-I seesaw model. 
 Here $\ell^{\alpha}_L$ denotes
the $SU(2)$ lepton doublet, 
as the EW symmetry is unbroken at this stage.}
\label{fig:HiggsLoops}
\end{figure}

 It has been recently suggested that the heavy Majorana states, $N_i$, 
introduced in the type I seesaw model could be responsible for the dynamical 
generation of the scalar potential of the Standard Model, 
in addition to that of neutrino masses~\cite{Brivio:2017dfq,Brivio:2018rzm}. 
In this scenario, dubbed the Neutrino Option, the classical potential is given by
\begin{equation}
\label{eq.classical_potential}
V_0(\Phi) = -\frac{M_{H0}^2}{2}\Phi^\dagger \Phi + \lambda_0 (\Phi^\dagger\Phi)^2,
\end{equation}
%
and is assumed to be nearly conformal at the seesaw scale: 
$M_{H0}(\mu\gtrsim M)\simeq 0$, $\lambda_0(\mu\gtrsim M)\neq 0$, 
$\mu$ being the renormalisation scale. This condition can be realized for instance in the conformal model proposed in~\cite{Brdar:2018vjq,Brdar:2018num}.
Radiative corrections to both $M_H^2$ and $\lambda$  are generated 
via
the diagrams of \figref{fig:HiggsLoops}, 
thus breaking scale invariance at the quantum level. 
%
At energies $\mu < M$ the $N_i$ fields can be integrated out and 
decoupled from the spectrum. One is then left with an 
Effective Field Theory (EFT) in which the leading seesaw contributions 
are encoded in the Weinberg operator for neutrino masses 
(stemming from the seesaw-EFT matching at tree-level) and 
in finite \emph{threshold matching contributions} to the Higgs 
potential parameters $\Delta M_H^2,\, \Delta\lambda$ 
(stemming from the one-loop matching).
The latter are identified as the zeroth order term in the 
$E/M$ expansion of the loops, computed with dimensional 
regularisation within the $\overline{MS}$ renormalisation scheme.
For the case of two heavy Majorana neutrinos 
with $M_2 = x_M M_1,\, x_M\geq1$ they read~\cite{Brivio:2018rzm}:
\begin{equation}
\begin{aligned}
\label{eq.Delta_mH}
\Delta M_H^2 &= \frac{M_1^2}{8\pi^2}\left(|Y_1|^2 + x_M^2 |Y_2|^2\right)\,,\\
\Delta\lambda & =
-\frac{1}{32\pi^2}\left[5|Y_1|^4 + 5|Y_2|^4 
+ 2 \,{\rm Re}(Y_1\cdot Y_2^*)^2\left(1-\frac{2\log x_M^2}{1-x_M}\right) \right.\\
&\left. + 2 \,{\rm Im}(Y_1\cdot Y_2^*)^2\left(1-\frac{2 \log x_M^2}{1+x_M}\right)\right]\,,
\end{aligned}
\end{equation}
%
with $Y_i$ the $i$th column of the 
matrix of neutrino Yukawa couplings.
In the limit $M_2=M_1$ ($x_M=1$) these reduce to 
\footnote{Note that although modified Feynman rules must be used when 
the heavy Majorana neutrinos are nearly degenerate in mass, this is only 
important when results, such as the CP asymmetry, depend on the difference 
of masses. In the calculation of the Higgs mass parameter, the contributions 
from each heavy Majorana neutrino are summed and any correction depending on 
the difference of the masses is negligible.}:
\begin{equation}
\begin{aligned}
\Delta M_H^2 &= \frac{M_1^2}{8\pi^2}\left(|Y_1|^2 + |Y_2|^2\right)\,,\\
\Delta\lambda &=
-\frac{5}{32\pi^2}\left(|Y_1|^4 + |Y_2|^4 + 2\,{\rm Re}(Y_1\cdot Y_2^*)^2\right)
-\frac{1}{16\pi^2}{\rm Im}(Y_1\cdot Y_2^*)^2\,.    
\end{aligned}
\end{equation}
%
The values of $m_{H},\lambda$ at the EW scale can be extrapolated 
using the renormalisation group equations (RGEs) of 
the SM~\cite{Buttazzo:2013uya} 
(as the heavy neutrinos are not present in the spectrum) 
with the following  boundary conditions: 
\begin{equation}
M_H^2(\mu=M_1) \equiv \Delta M_H^2 
\quad\text{ and }\quad
\lambda(\mu=M_1)\equiv\Delta\lambda + \lambda_0\,.    
\end{equation}
%
The condition on $M_H$ places the strongest constraint on the parameter 
space of the Neutrino Option, requiring $M_i\lesssim 10^7$~GeV and 
$|Y_{\alpha i}|\sim \unit[1]{TeV}/M_i$ 
(barring tunings between the Yukawa entries) in order to reproduce 
the correct Higgs mass and, at the same time, be consistent 
with the constraints from neutrino oscillation experiments. 
As neither $M_H$ nor the light neutrino masses change significantly under RGE running, this result is consistent to a good approximation with the order-of-magnitude estimate
\begin{equation}
M_H^2(\mu=v)\simeq \Delta M_H^2 \sim \frac{M_i^2|Y_{i}|^2}{8\pi^2}
\quad\text{ and }\quad
m_\nu(\mu=v) \simeq \frac{v^2|Y_{i}|^2}{2M_i}\,.
\end{equation}
Within this region of the parameter space the contribution to the Higgs quartic term is 
$|\Delta\lambda|\leq 10^{-7}$. As a consequence,   the threshold matching contribution 
is always negligible in comparison to the coupling in the 
classical potential which has to be positive and of loop size 
($\lambda_0\simeq 0.01-0.05$) in order to obtain the correct 
scalar potential at the EW scale~\cite{Brivio:2018rzm}.

%
\subsection{The framework of leptogenesis}
\label{subsec:LGframework}
%
%

 The type I seesaw provides a possible mechanism for the generation 
of the matter-antimatter asymmetry if the heavy Majorana neutrinos 
$N_i$ decay out-of-equilibrium, in
C-/CP- and lepton-number-violating processes in the early Universe. 
These decays produce a leptonic matter-antimatter asymmetry which may be 
transformed in part into the observed baryon asymmetry by SM sphaleron 
processes which violate the $B+L$ charge but conserve the $B-L$ one. 
Any such process which  generates a lepton asymmetry that is then converted into a 
baryon asymmetry is called leptogenesis.

The baryon-to-photon ratio, which parametrises the baryon asymmetry of the Universe (BAU),  is defined as
\begin{equation}
\eta_B \equiv \frac{n_B-n_{\overbar{B}}}{n_\gamma},
\end{equation}
%
where $n_B$, $n_{\overline{B}}$ and $n_\gamma$ are the number 
densities of baryons, antibaryons and photons respectively.
Its value has been inferred by two independent methods: 
the measurement from Big-Bang nucleosynthesis (BBN), which occurs 
when the temperature of the Universe satisfies 
$T \lesssim 1 \text{ MeV}$~\cite{Patrignani:2016xqp}, 
and the measurement of $\eta_B$ from 
Cosmic Microwave Background radiation (CMB) 
data~\cite{Ade:2015xua}, which probes it at the time of recombination, 
$T \lesssim 1 \text{ eV}$. As the latter value has been more precisely 
measured, we choose to use it  throughout:
\[
{\eta_{B}} _{\text{CMB}} = \left(6.02-6.18\right)\times 10^{-10},
\]
where the above is the $3\sigma$ range of the measurement. 

The Boltzmann equations for thermal leptogenesis 
provide a semi-classical description of the time evolution of 
the heavy neutrino densities, 
$n_{N_i}$ ($i \in \{1,2,3,\hdots\}$) 
with equilibrium distributions $n^\text{eq}_{N_i}$,
and the lepton asymmetry, $n_{B-L}$.
\footnote{All number densities are normalised to a volume containing 
a single heavy Majorana neutrino in ultra-relativistic thermal equilibrium.}
At the scales $T \lesssim 10^7$ GeV, at which  the Neutrino Option can be viable,
the flavour effects in leptogenesis are important.
 Each flavour component of the leptons produced in $N_i$ decays feels a different 
interaction with the plasma of the early Universe and decoheres from the other two. 
Consequently, the kinetic 
equations must separately describe the time evolution of the asymmetries in the tauon, muon and electron 
lepton charges,
$n_{\tau \tau}$, $n_{\mu \mu}$, $n_{e e}$. 
The total baryon asymmetry is computed by taking the sum 
$n_{B-L} = n_{\tau \tau} + n_{\mu \mu} + n_{e e}$ and 
 multiplying it
by a factor 
such that $\eta_B \approx 0.01 n_{B-L}$. In this regime all three lepton 
flavour asymmetries  $n_{\tau \tau}$, $n_{\mu \mu}$ and $n_{e e}$ evolve independently
and the  Boltzmann equations are given by:
\begin{equation}
\begin{aligned}
\label{eq:BE2F}
\frac{dn_{N_{i}}}{dz}&=-D_{i}(n^{}_{N_{i}}-n^\text{eq}_{N_{i}}),\\
\frac{dn_{\alpha \alpha}}{dz}& =\sum_i\left(\epsilon^{(i)}_{\alpha \alpha} D_{i}(n^{}_{N_{i}}-n^\text{eq}_{N_{i}})-p_{i \alpha}W_{i}n_{\alpha \alpha} \right)\,,~~i=1,2\,,~\alpha = e,\mu,\tau\,.
\end{aligned}
\end{equation}
where $z \equiv M_1/T$ and increases with time,  
$p_{i \alpha}$ and $\overline{p}_{i \alpha}$ are
the projection probabilities from decay state $i$ to the flavour state
$\alpha$ for particles and antiparticles, respectively, and 
$\epsilon^{(i)}_{\alpha \alpha}$ is the CP asymmetry in 
the lepton charge $\alpha$ generated in the decay of the heavy Majorana 
neutrino $N_i$. The projection probabilities are given by
\begin{equation}
\begin{aligned}
p_{i \alpha} \equiv \left| c_{i \alpha} \right|^2,\\
\overline{p}_{i \alpha} \equiv \left| \overline{c}_{i \alpha} \right|^2,
\label{eq:pia}
\end{aligned}
\end{equation}
where $c_{i \alpha}$ and $\overline{c}_{i \alpha}$ are the projection coefficients of the charged lepton products of the decay of the $i$th heavy Majorana neutrino ($N_i \rightarrow \overline{\Phi} l_i$, $N_i \rightarrow \Phi \overline{l}_i$) in
\[
|\ell_i \rangle = \sum_{\alpha} c_{i \alpha} | l_{\alpha} \rangle,
\]
\[
|\overline{\ell}_i \rangle = \sum_{\alpha} \overline{c}_{i \alpha} | \overline{l}_{\alpha} \rangle,
\]
%
%
and $\epsilon^{(1)}_{\alpha \alpha}$ the CP asymmetry (given in \equaref{eq:epsi}) with $i=1,2,3$ and ~$\alpha = e,\mu,\tau$. For further details we refer the reader to Ref. \cite{Blanchet:2011xq}.

The decay parameter, $D_i$,  describes the decay of 
$N_i$ and is defined in terms of the heavy neutrino decay rate 
$\Gamma_i \equiv \Gamma_i \left(N_i \rightarrow \overline{\Phi} \ell_{i} \right)$
with $\Phi$ the Higgs and the CP-conjugate rate, $\overline{\Gamma}_i \equiv \Gamma_i \left(N_i \rightarrow \Phi \overline{\ell}_{i} \right)$, and Hubble rate, $H$~\cite{Buchmuller:2004nz}:
\begin{equation}
D_i \equiv \frac{\Gamma_i+\overline{\Gamma}_i}{Hz}.
\end{equation}
%
Likewise, the washout factor is defined in terms of the heavy 
neutrino inverse decay rate $\Gamma^{\text{ID}}_i\equiv \Gamma_i \left(\overline{\Phi} \ell_{i} \rightarrow N_i \right)$ and the CP-conjugate 
inverse decay rate $\overline{\Gamma}^{\text{ID}}_i\equiv \Gamma_i \left(\Phi \overline{\ell}_{i} \rightarrow N_i \right)$ is
\begin{equation}
W_i \equiv \frac{1}{2}
\frac{\Gamma_i^{\text{ID}}+\overline{\Gamma}_i^{\text{ID}}}{Hz},
\end{equation}
%
We also define the washout-parameter 
\begin{equation}
K_i \equiv \frac{\tilde{m}_i}{m_*},
\quad
\text{where}
\quad
\tilde{m}_i \equiv \frac{\left( Y^\dagger Y \right)_{ii} v^2}{M_i},
\end{equation}
and $m_* = (16 \pi^2 v^2/ 3 M_{P}) \sqrt{(g_* \pi)/5} \approx 10^{-3}$ eV with $M_{P}$ the Planck mass.

As will be discussed in further detail in \secref{subsec:LGatNOscales},
we find it necessary to take the two heavy Majorana neutrinos to 
be nearly-degenerate in mass 
$\Delta M = M_2 - M_1 \ll M \equiv \frac{1}{2}(M_1 + M_2)$ 
 and thereby
$\Delta M \sim \Gamma$. 
In this case, one is concerned with 
resonant leptogenesis \cite{Pilaftsis:1997jf,Pilaftsis:2003gt} 
where the self-energy contribution to the CP asymmetry parameter 
may become large. Such enhancement of the asymmetry can be significant, allowing the energy scale for successful leptogenesis to be lowered by several orders of magnitude. For this reason, resonant leptogenesis has been most often explored in the literature within scenarios with Majorana masses of the order of a few TeV. Here we apply this paradigm to a wider energy range. The peculiarity of resonant leptogenesis is that non-negligible contributions to the CP-asymmetry can be induced by mixing and oscillation of the heavy Majorana neutrinos. The mixing effects  come 
from the possibility of off-diagonal transitions in the self-energy 
diagrams at $T=0$ which are included through  use of the resummed 
Yukawa couplings \cite{Dev:2014laa}. In the same regime, $\Delta M \sim \Gamma$, the thermal 
contributions to the self-energies are also important. This provides 
an extra contribution to the CP asymmetries in processes where 
on-shell heavy Majorana neutrinos oscillate in flavour space due to their 
interactions with a thermal background \cite{Dev:2014wsa}.  
The CP asymmetry which takes account of both the mixing and 
oscillation of the heavy Majorana neutrinos has the form
\cite{Bambhaniya:2016rbb}:
\begin{equation}
\epsilon^{(i)}_{\alpha \alpha} = \sum_{j \neq i} \frac{\operatorname{Im}\left[Y^\dagger_{i \alpha} Y_{\alpha j} \left(Y^\dagger Y \right)_{i j}\right]+\frac{M_{i}}{M_{j}} \operatorname{Im}\left[Y^\dagger_{i \alpha} Y_{\alpha j}\left(Y^\dagger Y\right)_{j i}\right]}{\left(Y^\dagger Y\right)_{i i}\left(Y^{\dagger} Y\right)_{j j}}\left(f_{i j}^{\operatorname{mix}}+f_{i j}^{\mathrm{osc}}\right)\,,
\label{eq:epsi}
\end{equation}
%
where
\begin{equation}
f_{i j}^{\operatorname{mix}} =\frac{\left(M_{i}^{2}-M_{j}^{2}\right) M_{i} \Gamma_{j}}{\left(M_{i}^{2}-M_{j}^{2}\right)^{2}+M_{i}^{2} \Gamma_{j}^{2}}\,,
\end{equation}
%
and
\begin{equation}
f_{i j}^{\mathrm{osc}} =\frac{\left(M_{i}^{2}-M_{j}^{2}\right) M_{i} \Gamma_{j}}{\left(M_{i}^{2}-M_{j}^{2}\right)^{2}+\left(M_{i} \Gamma_{i}+M_{j} \Gamma_{j}\right)^{2} \frac{\operatorname{det}\left[\operatorname{Re}\left(Y^{\dagger} Y\right)\right]}{\left(Y^{\dagger} Y\right)_{i i}\left(Y^{\dagger} Y\right)_{j j}}}\,.
\label{eq:fosc}
\end{equation}

%
\subsection{Leptogenesis at the scales required for the Neutrino Option}
\label{subsec:LGatNOscales}
%
%

The largest value of the heavy Majorana neutrino masses compatible with the 
Standard Model Higgs mass in the Neutrino Option scenario is 
$M_i \sim 10^7$ GeV. A lower bound on the heavy Majorana neutrino masses 
can be set by the requirement of perturbativity of the neutrino Yukawa couplings
and is $M_i \gtrsim \mathcal{O} \left( 10^{2} \right) $ GeV\footnote{
The lower bound on $M_{1}$ as shown in Fig.~(3) of \cite{Brivio:2018rzm}
comes from the additional assumption that $\left|\sin (x + i y) \right| < 1$, that constrains the width of the allowed bands in the figure.
This assumption was introduced in order to forbid explicitly fine tunings in the flavor space, but can be relaxed in full generality.
}.
In such a regime,  in order for 
 the neutrino masses to satisfy 
the existing limits, fine tuned cancellations must exist
between the tree level seesaw and the one-loop contributions to the light neutrino masses.

From the lower bound derived from perturbativity arguments to the upper bound from the
Neutrino Option itself, there are  two possible mechanisms of leptogenesis viable in the 
relevant heavy Majorana neutrino mass range:
\begin{enumerate}
 \item Thermal leptogenesis with enhanced $R$-matrices
~\cite{Moffat:2018wke} (see also 
\cite{Blanchet:2008pw,Antusch:2009gn}).
\item  Resonant leptogenesis with nearly degenerate heavy Majorana 
neutrino masses~\cite{Pilaftsis:2003gt,Pilaftsis:2005rv,Hambye:2001eu,Hambye:2004jf,Cirigliano:2006nu,Xing:2006ms,Branco:2006hz,Chun:2007vh,Kitabayashi:2007bs}.   
 \end{enumerate}

In this subsection, we shall argue that the fine-tuned scenario 
detailed in $(1)$ is incompatible with the Neutrino Option, and therefore 
justify our exclusive use the resonant method for the investigations 
in this work. The arguments we present here are valid 
in the more general case of three heavy Majorana neutrinos
which we consider below.
 
 In the fine-tuned case the elements of the $R$-matrix elements tend to be 
large and the one-loop contribution to the light neutrino masses should 
be incorporated through  modification of the Casas-Ibarra  parametrisation 
~\cite{Lopez-Pavon:2015cga}.
In this case, the $R$-matrix has the structure~\cite{Moffat:2018wke}:
\begin{equation}
\label{eq:RStructure}
R \approx
\left(
\begin{array}{ccc}
 R_{11} & R_{12} & R_{13} \\
 \pm i R_{22} & R_{22} & R_{23} \\
 -R_{22} & \pm i R_{22} & \pm i R_{23} \\
\end{array}
\right)\,,
\end{equation}
%
Here $|R_{22}| \gg |R_{1i}|, |R_{23}|$ for $i \in \{1,2,3\}$, which results,
in \cite{Moffat:2018wke}
\begin{equation}\label{eq:fine-tuning}
m^{\text{tree}} \approx - m^{\text{1-loop}}\,.
\end{equation}
A typical fine-tuned leptogenesis solution, which allows for partial 
cancellations between the tree and one-loop level contributions to the 
light neutrino masses, requires $M_1 \approx 5 \times 10^{6}$ GeV 
(see the scenarios of ~\cite{Moffat:2018wke}) and correspond to large 
values $\sqrt{\Delta M_H^2} \sim 8 \times 10^6$ GeV. The latter value owes its magnitude to the dependence of $\Delta M_H^2$ on the $R$-matrix elements which are themselves large in order to provide the fine-tuning of \equaref{eq:fine-tuning}. 

We  performed  a numerical search of the parameter space with heavy neutrino masses  
$\sim10^6$ GeV and found no points which simultaneously satisfy the 
requirements of the Neutrino Option and leptogenesis even when we allowed
$m^{\text{tree}}$ and $m^{\text{1-loop}}$ to cancel to $\sim 0.1\%$. When no constraint was placed on the levels of fine-tuning it was possible to find solutions with the desired value of $\eta_B$ and 
$\sqrt{\Delta M_H^2} \sim 100$ GeV. Such solutions corresponds to an  $R$-matrix with very large entries, $\lvert R_{ij}\rvert \sim 10^{12}$, and very small (physically unreasonable) heavy neutrino masses $M_i \sim 10^{-2}$ GeV. The fine-tuned cancellation between the tree- and one-loop light neutrino masses is so complete that the higher-order radiative corrections to the light neutrino masses dominate and exceed the light neutrino mass bound. Using the estimate that the two-loop contribution to the light neutrino masses is
\[
\left| m^\text{two-loop} \right| \sim \frac{1}{16 \pi^2} \left| m^{\text{one-loop}} \right| \text{max.}(|Y|)^2 \approx \frac{1}{16 \pi^2} \left| m^{\text{tree}} \right| \text{max.}(|Y|)^2,
\]
where $\text{max.}(|Y|)$ is the largest element of the matrix of absolute values of neutrino Yukawa couplings, and where we use the approximate equality of \equaref{eq:fine-tuning}, we estimate
\[
m^\text{two-loop} \sim 10^3 \text{ GeV},
\]
which is  well-excluded by the experimental constraints. In summary, 
we found only physically 
non-viable
solutions involving fine-tuned non-resonant leptogenesis. 
For this reason, for the remainder of this work we restrict ourselves 
to resonant leptogenesis in which $M_1 \approx M_2$ 
and where $N_3$ is decoupled.

%
%
\section{Results}
\label{sec:results}
In this section we present our main results. In \secref{sec:lowerbound}, we derive the lower bound on the mass scale $M$ for which leptogenesis is viable within the Neutrino Option and explore the available parameter space and in \secref{sec:upperbound} we find the corresponding upper bound. In \secref{sec:LGLECPVNO} we find the conditions under which it is possible to achieve successful leptogenesis in the Neutrino Option when the CP violation comes purely from the low-energy phases of the PMNS matrix.
%
\subsection{Lower bound on the heavy Majorana neutrino masses}\label{sec:lowerbound}
%
%
 In this section we determine the range of  heavy Majorana neutrino masses
in which both the Neutrino Option and leptogenesis are viable. 
We shall always work in the nearly degenerate case 
$\Delta M \equiv M_2-M_1 \ll M$ where $M \equiv (M_1+M_2)/2$. 
Evaluating $\Delta M_H^2$ at $\mu \sim M$
as is calculated in~\cite{Brivio:2018rzm}, the threshold correction is
\begin{equation}
\Delta M_H^2 = \frac{1}{8 \pi^2} \text{Tr}\left[Y M^2 Y^\dagger \right].
\end{equation}
%
By substitution of the Casas-Ibarra parametrisation of 
\equaref{eq:CINO} or \equaref{eq:CIIO}, this becomes
\begin{equation}
\label{eq:HiggsMassCI}
\Delta M_H^2 = \frac{1}{8 \pi^2 v^2} \cosh{\left( 2 y \right)} M^3 \left( m_1 + m_2 + m_3 \right),
\end{equation}
%
where $m_1 = 0$ ($m_3=0$) for NO (IO) spectrum and the neutrino parameters run with
the scale M. This effect amounts to a few percent for the light neutrino masses and is implemented here using the RGEs of Refs.~\cite{Casas:1999tg,Antusch:2003kp}. The remainder of the neutrino parameters change less significantly when RG evolved, so their scale dependence will be neglected.
Note that the $x$ parameter (the real part of $\theta$ 
which parametrises the $R$ matrix), cancels in~\eqref{eq:HiggsMassCI} due to the near-diagonality of 
the heavy Majorana mass matrix. The Neutrino Option is then satisfied if 
the Standard Model $\overline{MS}$ Higgs mass, when renormalisation group 
evolved to the scale $M$, matches the values given by $\Delta M_H^2$. 
The running of the SM parameters is taken into account implementing the RGE of Ref.~\cite{Buttazzo:2013uya} to the highest available accuracy and with the numerical inputs reported in Table~\ref{tab:SMinputs}. 

\begin{table}[t]
\centering
\renewcommand{\arraystretch}{1.2}
\begin{tabular}{lclc}
 \toprule
inputs & value (GeV)\\
\specialrule{2.5pt}{1pt}{1pt}
$v$         & 174.10\\
$M_H $      & 125.09 \\
$M_t $     & 173.2\\
\\
\bottomrule
\end{tabular}
\hspace*{4mm}
\begin{tabular}{lclc}
 \toprule
\multicolumn{4}{l}{RGE boundary conditions at $\mu=M_t$}\\
\specialrule{2.5pt}{1pt}{1pt}
$ \lambda$&       0.1258   &$ M_H (\text{GeV})$&     131.431\\
$ g_1$&           0.461    &$ Y_t$&         		  0.933  \\
$ g_2$&           0.644      &$ Y_b$&        		   0.024   \\
$ g_3$&           1.22029   &$ Y_\tau$&       	 0.0102 
  \\
  \bottomrule
\end{tabular}
\caption{Values of the relevant SM parameters adopted in the numerical analysis, consistent with Ref.~\cite{Brivio:2018rzm}. The RGE boundary conditions are computed at the highest accuracy provided in Ref.~\cite{Buttazzo:2013uya}. \\
}
\label{tab:SMinputs}
\end{table}

The lower bound on $M$ for leptogenesis in the Neutrino Option is the lowest 
scale for which the correct baryon asymmetry results whilst still satisfying 
\equaref{eq:HiggsMassCI}. We apply the approximate analytical solution
\begin{equation}
\label{eq:nB-LApprox}
\begin{aligned}
n_{B-L} & \approx \frac{\pi^2}{6 z_d} n^\text{eq} \left( 0 \right) 
\sum_{\alpha = e}^\tau \frac{\epsilon^{(1)}_{\alpha \alpha}}{K_1 p_{1 \alpha}},
\end{aligned}
\end{equation}
%
in the derivation of which we have eliminated $\epsilon^{(2)}_{\alpha \alpha}$ with the approximation $\epsilon^{(2)}_{\alpha\alpha}/K_2 p_{2 \alpha} \approx \epsilon^{(1)}_{\alpha \alpha}/K_1 p_{1 \alpha}$. For the lowest heavy Majorana neutrino mass scale $M$, the value of $y$ 
(the imaginary part of $\theta$) will be largest as can be seen 
from inspection of \equaref{eq:HiggsMassCI}. Approximating the terms in the sum under 
the assumption that $e^y \gg e^{-y}$ 
 (to be justified later),
we find
\begin{equation}
\label{eq:EfficiencyNOIO}
\begin{aligned}
\frac{\epsilon^{(1)}_{\alpha \alpha}}{K_1 p_{1 \alpha}} & \approx 16 m_{*} \left( f_{\text{osc}} + f_{\text{mix}} \right) \frac{m_2 - m_3}{\left(m_2+m_3\right)^2} e^{- 4 y} \sin 2x\,, \quad \text{Normal Ordering\,,} \\
\frac{\epsilon^{(1)}_{\alpha \alpha}}{K_1 p_{1 \alpha}} & \approx 16 m_{*} \left( f_{\text{osc}} + f_{\text{mix}} \right) \frac{m_1 - m_2}
{\left(m_1+m_2\right)^2} e^{- 4 y} \sin 2x\,, \quad \text{Inverted Ordering}\,.
\end{aligned}
\end{equation}
Using the same approximation, the factor 
$\propto \text{det} [\text{Re}(Y^\dagger Y)]$
in the denominator of $f_{\text{osc}}$ 
(see Eq. (\ref{eq:fosc}))
becomes
\[
\frac{\text{det} \left[ \text{Re} \left( Y^\dagger Y \right) \right]}{ \left( Y^\dagger Y \right)_{11} \left( Y^\dagger Y \right)_{22} } \approx 1.
\]
From \equaref{eq:EfficiencyNOIO} we observe that the parameter $y$ 
exponentially suppresses the final asymmetry $\eta_B$ while enhancing 
the Higgs mass of \equaref{eq:HiggsMassCI}. 
To the level of accuracy in the 
approximation, the contribution of each flavour is identical. 
Although the   left-hand side carries a flavour index $\alpha$, 
the right-hand side is independent of this index. As the flavour information 
is contained in the PMNS elements, we expect that at large $y$, the solutions 
for successful leptogenesis in the Neutrino Option should have only a weak 
dependence on the PMNS phases 
(in terms that have been neglected in \equaref{eq:EfficiencyNOIO}).

\begin{figure}[t]
    \centering
    \includegraphics[width=0.7\textwidth,center]{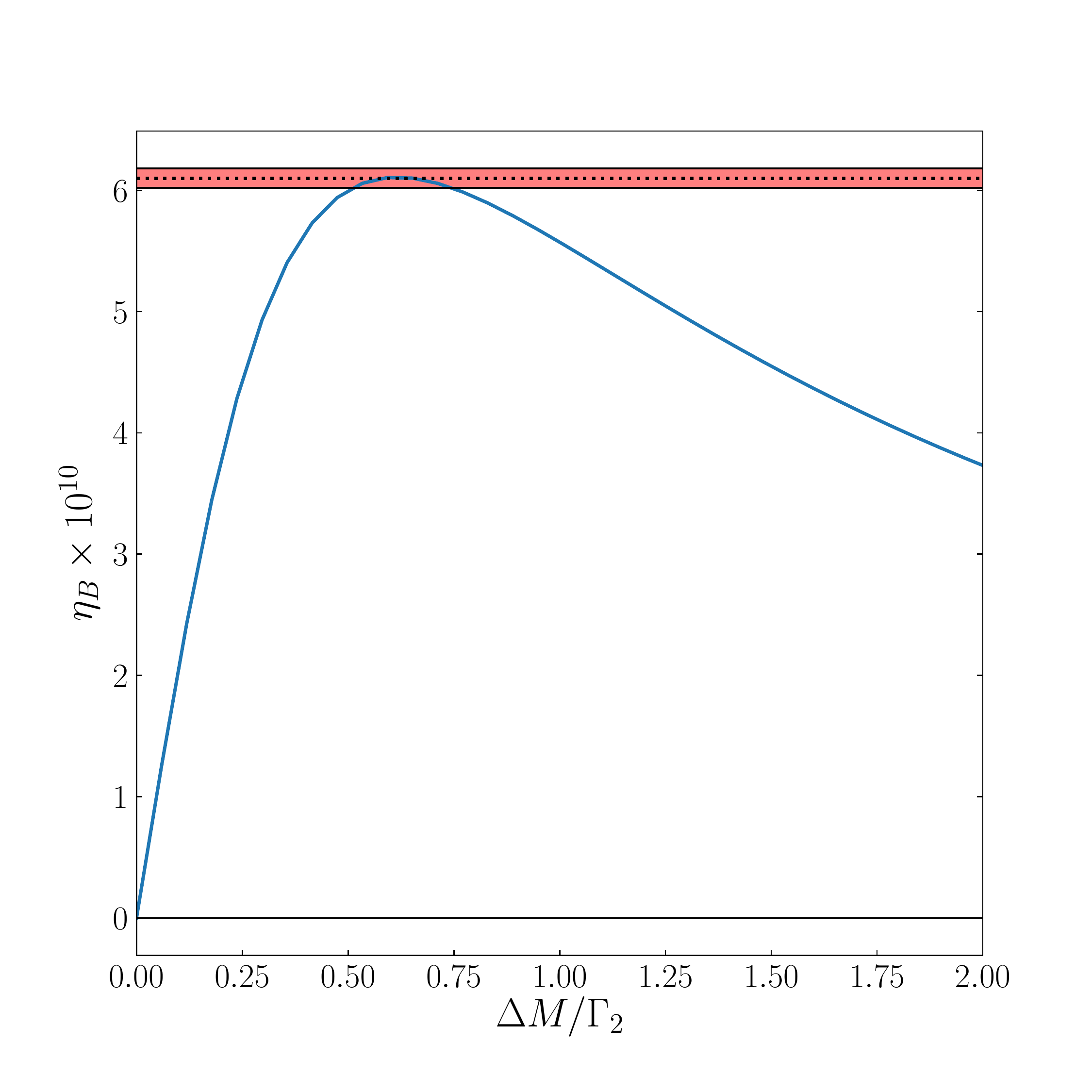}
    \caption{The baryon asymmetry as a function of the heavy Majorana 
neutrino mass splitting divided by the decay rate of $N_2$ at the lower-bound 
of $M$
for successful leptogenesis in the Neutrino Option. The plots take an 
identical form for both normal ordering 
(for which $\Gamma_2 = 1.62 \times 10^{-2}$ GeV) and inverted ordering 
(for which $\Gamma_2 = 8.63 \times 10^{-3}$ GeV).}
    \label{fig:DeltaMEtaB}
\end{figure}
%

\begin{figure}[t]
    \centering
    \includegraphics{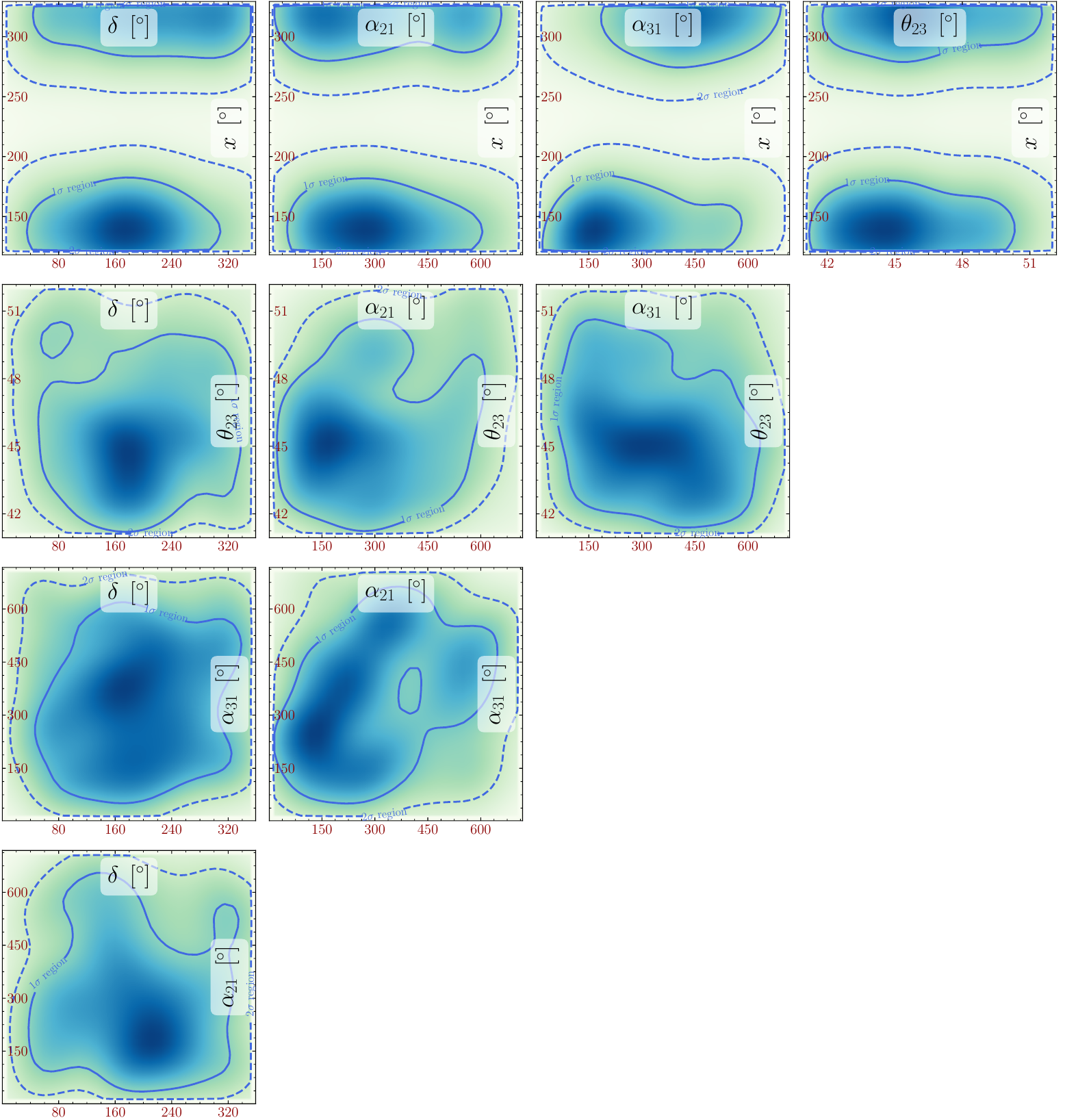}
    \caption{The triangle plot shows regions of the model parameter space compatible with the measured 	$\eta_{B}$
within $1\sigma$ and $2\sigma$
		for a normally ordered mass spectrum using resonant leptogenesis with $M = 1.2\times10^{6}$ GeV. The fixed parameters were set to be  $y=190.22^\circ$, $\theta_{12}=33.63^\circ$ and $\theta_{13} = 8.51^\circ$.}
    \label{fig:NOtriangle}
\end{figure}

The lower bound on $M$ may be found by maximising the terms in 
Eq. (\ref{eq:EfficiencyNOIO}) with respect to all parameters except $y$ and 
then finding the largest value of $y$ for which leptogenesis may be 
successful. 
From \equaref{eq:HiggsMassCI}, we observe the 
 scale of $M$ is determined from $y$ 
 (recalling that the light neutrino 
masses are to be run to the scale $M$) and may therefore infer the 
lower bound for successful leptogenesis in the Neutrino Option. 
The maximisation of the right-hand side of 
\equaref{eq:EfficiencyNOIO} occurs 
when $x = 135^\circ$ or $315^\circ$ and 
$\Delta M / \Gamma_2 \approx 0.61$ as shown in \figref{fig:NOtriangle}. We find that the values of $y$ that 
give agreement with $\eta_{B_{CMB}}$ are $y = 190.22^\circ$ for normal 
ordering and $y = 118.21^\circ$ for inverted ordering. 
These imply lower bounds for viable leptogenesis in the Neutrino Option of
\begin{equation*}
\begin{aligned}
& M > 1.2 \times 10^6 \text{ GeV} \quad \text{Normal Ordering,} \\
& M > 2.4 \times 10^6 \text{ GeV} \quad \text{Inverted Ordering.}
\end{aligned}
\end{equation*}
%
The difference in values for the two bounds is entirely determined by the 
difference in the factors $(m_2-m_3)/(m_2+m_3)^2$ and $(m_1-m_2)/(m_1+m_2)^2$ 
appearing for normal and inverted ordering respectively.
 We emphasise 
that the suppression factor $e^{-4y}$ occurring in 
\equaref{eq:EfficiencyNOIO} is sufficiently strong that the lower bounds are 
not strongly affected by the running of the parameters 
(although the bounds stated include the running of all SM parameters 
and the light neutrino masses).
%

As a cross-check we have also determined the lower bound on $M$ for which 
leptogenesis is viable within the Neutrino Option by numerically solving 
the resonant Boltzmann equations (we stress we solve the  Boltzmann equations 
and not the analytically approximated equations of \equaref{eq:EfficiencyNOIO})
for both normal and inverted ordering and scanned the available parameter 
space for $\eta_B = 6.1 \times 10^{-10}$.
We performed the parameter space exploration using {\sc
Multinest} \cite{Feroz:2007kg,Feroz:2008xx,Feroz:2013hea}  for a fixed scale $M$ but 
varied the splitting $\Delta M$, with $y$ fixed to the value that satisfies 
$M_H^2 \left(M_1 \right) = \Delta M_H^2 \left( M_1 \right)$. That is, 
a value of $M_1$ was chosen and $y$ was fixed to make the Neutrino Option work,
then $\delta, \alpha_{21}, \alpha_{31},x$ and $M_2$ were varied 
(none of which can spoil the generation of the Higgs potential 
once $M_1$ and $y$ are determined, provided $M_2$ does not differ 
significantly from $M_1$). We started at the maximum $M_1$ which was 
allowed by the Neutrino Option (which occurs when $y = 0^\circ$) and lowered it in small increments, 
performing a new parameter scan   at each of the successively smaller 
values of $M_1$. This procedure was stopped when the search
no longer yielded 
 points in the parameter space where leptogenesis was successful.
The lowest value of $M_1$ for which leptogenesis was viable was taken 
as our lower bound.

\begin{figure}[t]
    \centering
    \includegraphics{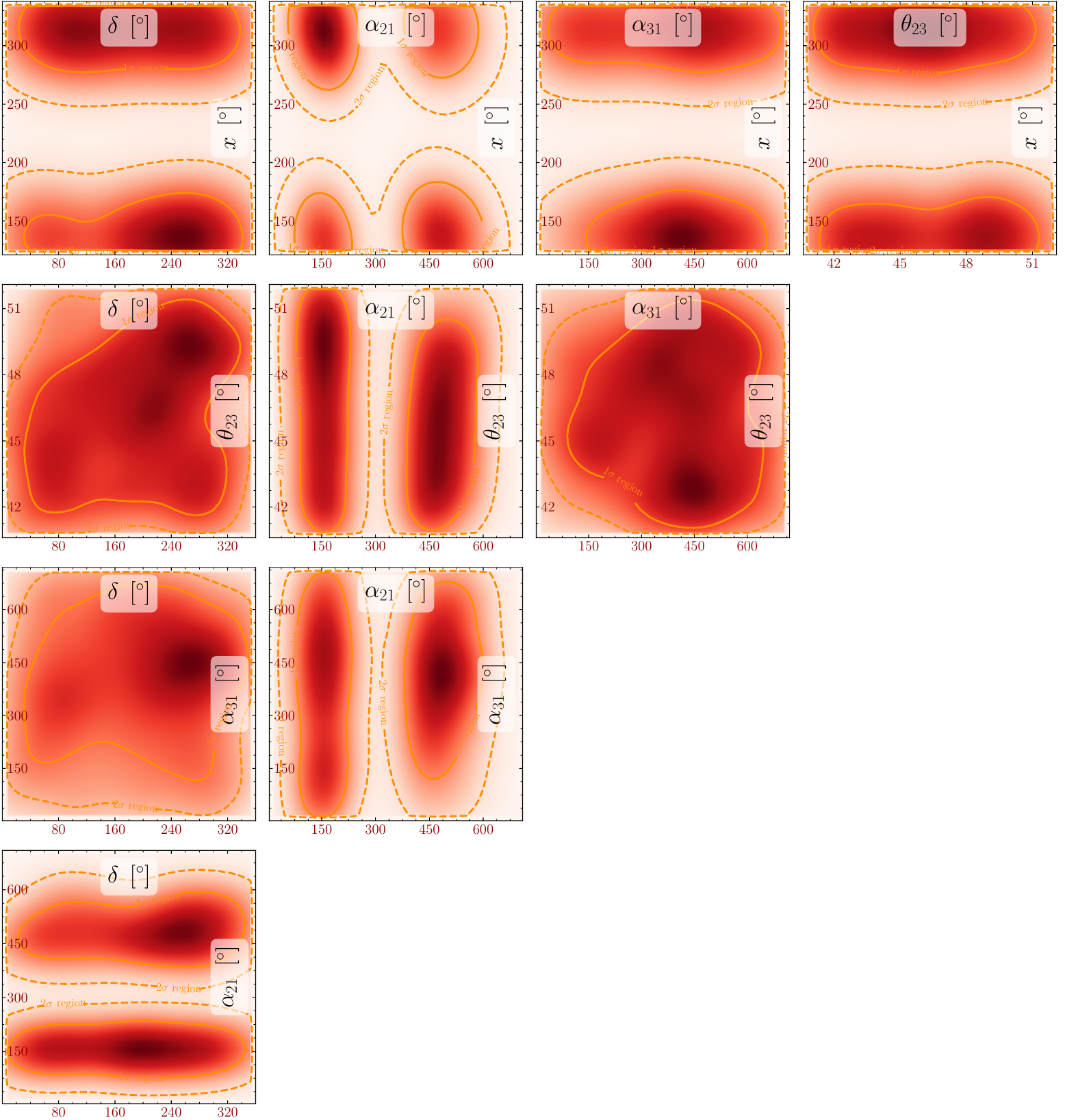}
    \caption{The triangle plot shows regions of the model parameter space compatible with the measured $\eta_{B}$ within one and two $\sigma$ for an inverted ordered mass spectrum using resonant leptogenesis with $M = 2.4 \times 10^{6}$ GeV. The fixed parameters were set to be  $y=118.21^\circ$, $\theta_{12}=33.63^\circ$ and $\theta_{13} = 8.51^\circ$.}
    \label{fig:IOtriangle}
\end{figure}

At the lowest successful value, for both normal and inverted ordering, 
we found that the numerical results, as shown in \figref{fig:NOtriangle} 
and \figref{fig:IOtriangle} (both of which are placed at lower bound on $M$), are in broad agreement with the statements made 
above based on the analytical approximations. 

In the inverted ordered case 
shown in \figref{fig:IOtriangle}, we can see a dependence on 
$\alpha_{21}$ that is
not accounted for in the approximated analytical expressions
and that 
is not present in the normal ordered case of \figref{fig:NOtriangle}. 
The reason is that the suppression factor $e^{- 4 y}$ is $\mathcal{O}(100)$ times smaller for inverted 
ordering, rendering the approximations of \equaref{eq:EfficiencyNOIO} 
slightly less accurate than for normal ordering. Terms which were neglected 
and which depend on $\alpha_{21}$,
contribute to a slight dip in the value of 
$\eta_B$ around $\alpha_{21} \sim 300^\circ$. The approximate independence 
from $\delta$ is preserved because terms in $\delta$ are multiplied by the 
relatively small factor $s_{13}$. Finally, for inverted ordering, 
independence of $\alpha_{31}$ is exact as it does not appear 
in the Yukawa matrix when $m_3 = 0$.

Finally, we note that, at the lower bound for both normal ordering and 
inverted ordering, the heavy Majorana neutrinos form a pseudo-Dirac pair 
\cite{Wolfenstein:1981kw,Petcov:1982ya} since, with $x = 135^\circ$,
\[
Y = U
\left(
\begin{array}{cc}
 0 & 0 \\
-\frac{\sqrt{M_1 m_2}}{\sqrt{2} v} \left( \cosh y + i \sinh y \right)& -\frac{\sqrt{M_2 m_2}}{\sqrt{2} v} \left( \cosh y -i \sinh y \right)\\
\frac{\sqrt{M_1 m_3}}{\sqrt{2} v} \left( \cosh y -i \sinh y \right) & -\frac{\sqrt{M_2 m_3}}{\sqrt{2} v} \left( \cosh y +i \sinh y \right) \\
\end{array}
\right),
\]
for which $Y_1 \approx i Y_2$ when $M_1 \approx M_2$.
At the lower bound where $\Delta M \ll M$, as the heavy Majorana neutrino mass matrix is diagonal, this condition implies that the CP phases of 
$N_1$ and $N_2$ are approximately opposite (see also \secref{sec:LGLECPVNO}) such that they form a Dirac pair when $M_1 = M_2$. Solutions of this kind may be motivated by assuming an approximate lepton number symmetry 
~\cite{Mohapatra:1986aw,Mohapatra:1986bd,Bernabeu:1987gr,Pilaftsis:1991ug,Ilakovac:1994kj,Akhmedov:1995ip,Akhmedov:1995vm,Abada:2010ym,Abada:2011hm,Alonso:2016onw,Gavela:2009cd,Dias:2011sq,Bazzocchi:2010dt,Ma:2009gu}.\\

%
\subsection{Upper bound on the heavy Majorana neutrino masses}\label{sec:upperbound}
%
%
The upper bound on $M$ for the Neutrino Option occurs when $y=0^\circ$ (\equaref{eq:HiggsMassCI}) . 
In this case one finds for normal ordering
\begin{equation}
\label{eq:EfficiencyNOUpperBounds}
\begin{aligned}
\frac{\epsilon^{(1)}_{\alpha \alpha}}{K_1 p_{1 \alpha}} & \approx  \frac{i m_{*} (m_2 - m_3) \sqrt{m_2 m_3} \cos x \sin x \left(U_{\alpha 2}^* U_{\alpha 3} - U_{\alpha 2} U_{\alpha 3}^* \right) \left( f_{\text{osc}} + f_{\text{mix}} \right) }{\left( m_2 \cos^2 x + m_3 \sin^2 x \right) \left( m_3 \cos^2 x + m_2 \sin^2 x \right) \left| \sqrt{m_2} \cos x U_{\alpha 2} + \sqrt{m_3} \sin x U_{\alpha 3} \right|^2} ,
\end{aligned}
\end{equation}
%
and for inverted ordering
\begin{equation}\label{eq:EfficiencyNOUpperBounds}
\begin{aligned}
\frac{\epsilon^{(1)}_{\alpha \alpha}}{K_1 p_{1 \alpha}} & \approx  \frac{i m_{*} (m_1 - m_2) \sqrt{m_1 m_2} \cos x \sin x \left(U_{\alpha 1}^* U_{\alpha 2} - U_{\alpha 1} U_{\alpha 2}^* \right) \left( f_{\text{osc}} + f_{\text{mix}} \right) }{\left( m_2 \cos^2 x + m_1 \sin^2 x \right) \left( m_1 \cos^2 x + m_2 \sin^2 x \right) \left| \sqrt{m_1} \cos x U_{\alpha 1} + \sqrt{m_2} \sin x U_{\alpha 2} \right|^2} .
\end{aligned}
\end{equation}
%
Unlike for the lower bound where these terms had a maximum value that 
was largely independent of the PMNS phases, here we find a strong dependence 
upon these low energy phases and apparently unrestricted enhancement factors. 
Thus, leptogenesis must be successful at the upper bounds of the 
Neutrino Option. Combining the lower bounds from requiring the Neutrino 
Option and leptogenesis to be viable simultaneously with the upper bounds 
from the Neutrino Option alone results in\footnote{
The upper bounds quoted here are slightly different from the one reported in Ref.~\cite{Brivio:2018rzm} owing to the running of light neutrino masses having been neglected in the latter. Numerically the difference amounts to about $\sim 5\%$ and has therefore limited significance.
}
\begin{equation*}
\begin{aligned}
&1.2 \times 10^6 < M \text{ (GeV)} < 8.8 \times 10^6 \quad \text{Normal Ordering,}\\
&2.4 \times 10^6 < M \text{ (GeV)} < 7.4 \times 10^6 \quad \text{Inverted Ordering.}
\end{aligned}
\end{equation*}
In this case the approximate lepton number symmetry need not be so precise
as it was for the lower bound.

\subsection{Leptogenesis from purely low energy CP violation in the Neutrino Option}
\label{sec:LGLECPVNO}

At the upper bound, for which $y=0^\circ$ and therefore $R$ is real, 
all CP violation comes from the phases of the PMNS matrix.
In this section we consider the possibility that the low energy CP phases 
($\delta$, $\alpha_{21}$ and $\alpha_{31}$) provide 
all of the CP violation necessary for leptogenesis to produce the observed BAU.
In past work it has been 
shown that these phases are sufficient sources of CP violation in a mass 
range $10^6 < M \text{ (GeV)} < 10^{13}$~\cite{Moffat:2018smo,Pascoli:2006ie,Pascoli:2006ci,Blanchet:2006be,Branco:2006ce,Anisimov:2007mw,Molinaro:2008rg,Molinaro:2008cw,Dolan:2018qpy}
(for a review see, e.g., \cite{Hagedorn:2017wjy}), where there must be some 
fine-tuning for heavy Majorana neutrino masses $M \lesssim 10^9$ GeV, 
if they are mildly hierarchical. Here we ask:
\textit{Is it possible for purely the low energy phases to provide 
CP violation for successful leptogenesis within the Neutrino Option?}

In the type I seesaw, both the light ($\nu_i$) and the heavy ($N_i$) 
neutrino mass states are Majorana and so satisfy the conditions:
\begin{equation}
\begin{aligned}
C \overline{\nu}_i^T & = \nu_i,\\
C \overline{N}_i^T & = N_i,
\end{aligned}
\end{equation}
%
where $C$ denotes the charge conjugation matrix. 

\begin{figure}[t]
\includegraphics[width=1.2\textwidth,center]{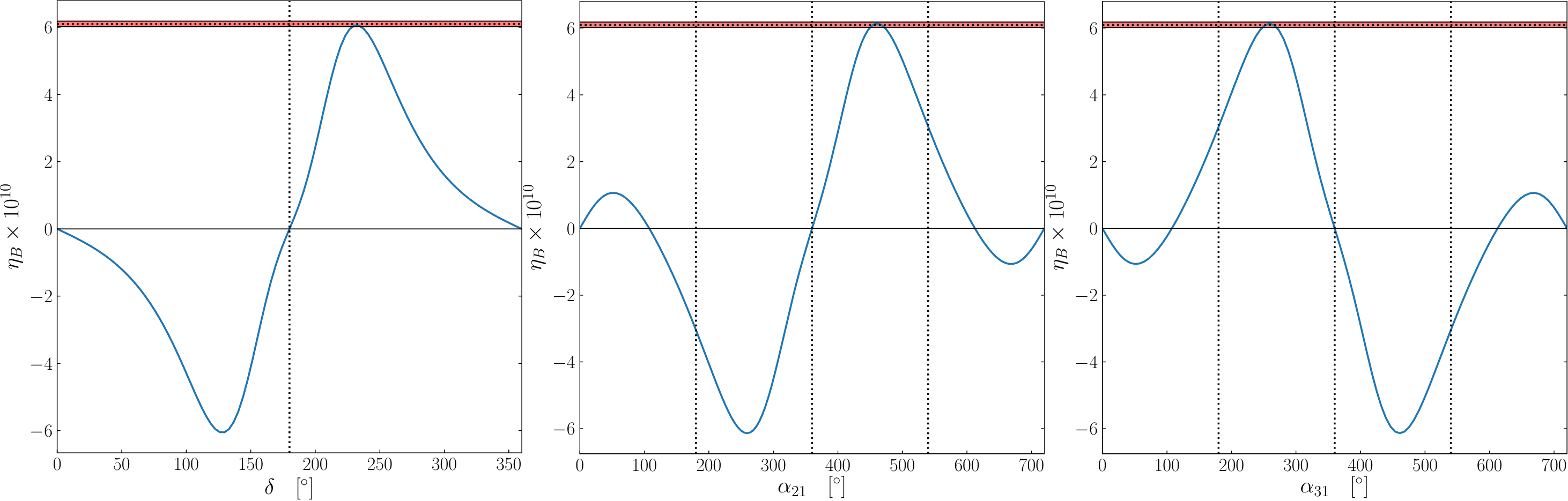}
\caption{The baryon asymmetry as a function of one of the 
CPV phases of the PMNS matrix $\delta$, $\alpha_{21}$ 
and $\alpha_{31}$ (the other two being set to zero),
with
the varied phase providing all the CP violation 
in leptogenesis, in the case of neutrino mass 
spectrum with normal ordering.
Vertical dotted lines represent values for which there is a complete 
CP symmetry in the neutrino sector.
See text for further details.} 
\label{fig:CPVNO}
\end{figure}

\begin{figure}[h!]
\includegraphics[width=\textwidth,center]{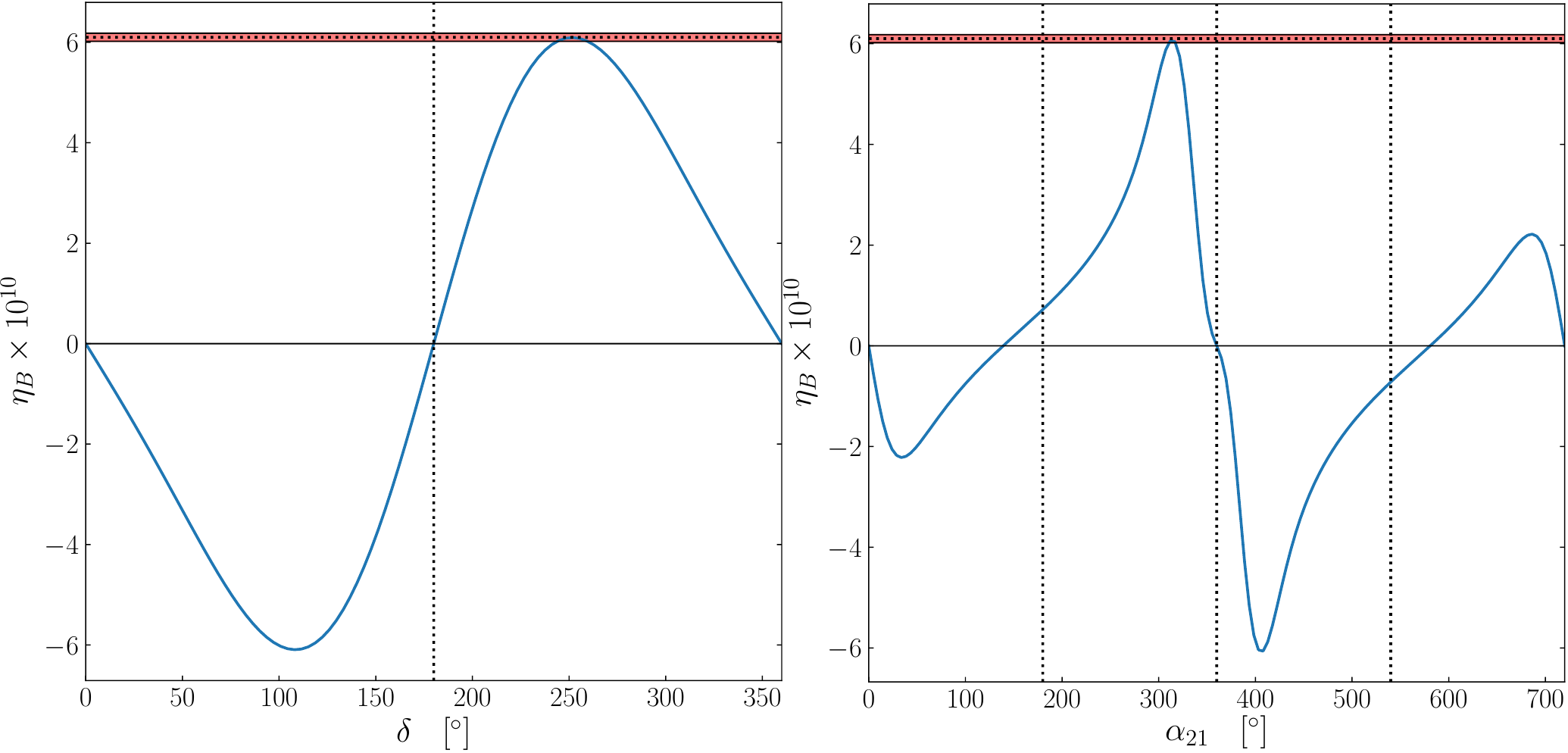}
\caption{
The same as in  \figref{fig:CPVNO} but 
for neutrino mass spectrum with inverted ordering.
The phase $\alpha_{31}$ decouples in this scenario and so
 a figure with the dependence of the baryon asymmetry on 
$\alpha_{31}$
is not shown.
See text for further details.}
\label{fig:CPVIO}
\end{figure}

In the case of CP invariance,
the Majorana fields  $N_i(x)$ and  $\nu_i(x)$ transform as follows 
under the operation of CP-conjugation
(see, e.g., \cite{Bilenky:1987ty}):
\begin{equation}
\begin{aligned}
U^{}_{CP} N_i\left(x\right) U_{CP}^{\dagger} & = i \rho^N_i \gamma_0 N_i\left(x'\right), \\
U^{}_{CP} \nu_i\left(x\right) U_{CP}^{\dagger} & = 
i \rho^{\nu}_i \gamma_0 \nu_i\left(x'\right),
\end{aligned}
\end{equation}
%
where $U_{CP}$ is the CP-conjugation operator,
$x'$ is the parity-transformed coordinate and 
$i\rho^{N}_i = \pm i$ and  $i\rho^{\nu}_i = \pm i$ are  
the CP parities of the respective Majorana fields. The conditions for CP invariance impose the 
following restrictions on the elements of 
the matrix of neutrino Yukawa couplings: 
\begin{equation}
\label{eq:YCPconditions}
Y_{\alpha i}^{*} = Y_{\alpha i} \rho^{N}_i\,,
\end{equation}
%
where the  unphysical phases in 
the CP transformations of the lepton and Higgs doublets have been set
to unity. The CP invariance condition imposed on the PMNS matrix gives the following relation \cite{Bilenky:1987ty}:
\begin{equation}
\label{eq:UCPconditions}
U^*_{\alpha j} = U_{\alpha j} \rho^{\nu}_j\,,~j \in \{1,2,3\},~
\alpha \in \{e,\mu,\tau\}\,.
\end{equation}
%
From the parametrisation of the Yukawa matrix this imposes 
the following conditions on the elements of the
$R$-matrix~\cite{Pascoli:2006ci}:
\begin{equation}
\label{eq:UCPRCPconditions}
 R^*_{ij} = R_{ij} \rho^{N}_i \rho^{\nu}_j\,,~~i,j \in \{1,2,3\}\,.
\end{equation}
%
In  \figref{fig:CPVNO} we show a scenario in which the Neutrino Option 
is satisfied at the upper bound of the normal ordered case and CP violation 
is provided entirely by the low-energy phases 
(each phase being zero unless varied). Similarly for inverted ordering 
of the light neutrinos we have the results depicted in Fig. 6.
The non-zero values of the baryon asymmetry at the CP conserving values 
$180^{\circ}$ and $540^{\circ}$ of $\alpha_{21}$ and $\alpha_{31}$
(of $\alpha_{21}$) 
seen in  \figref{fig:CPVNO} (in  \figref{fig:CPVIO})
are due to the CP violating interplay of the 
CP conserving PMNS and $R$ matrices~\cite{Pascoli:2006ci}. 
This is because there may be a choice of $\rho^{N}_j$ that satisfies the condition of \equaref{eq:UCPconditions}
and $\rho^{N}_i$ and $\rho^{\nu}_j$ which satisfy the condition  of \equaref{eq:UCPRCPconditions}  but there is no such combination of these phases  that are simultaneously satisfied at these values of  $\alpha_{21}$ and $\alpha_{31}$. 
As a result, the condition of  \equaref{eq:YCPconditions} on the Yukawa matrix is not fulfilled, leaving it a matrix of complex numbers (with elements not purely real or imaginary) corresponding to the generic CP-violating case.

Although only specific values of the Dirac and Majorana phases are compatible with successful leptogenesis in the context of the Neutrino Option, it is important to emphasise that only the Dirac is measurable at neutrino oscillation experiments. 
In the scenario the Majorana phases are the sole CP-violating phases, a CP-conserving measurement of $\delta$ would still be compatible with 
producing the BAU in this framework. 
Nonetheless, for each scenario (be it sole CP-violation from $\delta$, $\alpha_{21}$ or $\alpha_{31}$) a prediction for the value of the effective Majorana mass, $\lvert m_{ee}\rvert$, and the corresponding half-life of neutrinoless double beta decay, can be calculated. In principle, assuming precise knowledge of the neutrino masses and an accurate evaluation of nuclear matrix elements, this prediction can be tested by neutrinoless double beta decay experiments which will probe the inverted ordering regime in the near future  (see Ref. \cite{Pascoli:2005zb} for further discussion on this matter).


%
\section{Conclusions}
\label{sec:conclusions}
%
%
The aim of this paper is to examine resonant leptogenesis in the context of the Neutrino Option thereby taking one further step in constructing a self-consistent theory which simultaneously explains light neutrino masses, the predominance of matter over anti-matter, and the electroweak scale. 

We found that the  viable parameter space which can satisfy the Neutrino Option and leptogenesis are in the ranges $1.2 \times 10^6 < M \text{ (GeV)} < 8.8 \times 10^6$ and $2.4 \times 10^6 < M \text{ (GeV)} < 7.4 \times 10^6$ for normal and inverted ordering respectively, with successful leptogenesis requiring a pseudo-Dirac pair with masses such that $\Delta M/M \equiv (M_2 - M_1)/M \sim 10^{-8}$.  Interestingly, viable solutions for Neutrino Option leptogenesis allows for $\theta_{23}$ be to in the lower or upper octant (at the 2 $\sigma$ level) however for normal ordering there is a slight preference for solutions in the upper octant. In particular, we found that, generally, there is only a weak dependence on the low energy phases of the PMNS matrix $\delta$, $\alpha_{21}$ and $\alpha_{31}$ at the lower bounds on the viable mass range for both normal and inverted ordering (see \figref{fig:NOtriangle} and \figref{fig:IOtriangle}). The minor exception to this is $\alpha_{21}$ in the case of inverted ordering which must be approximately in the range $[90^\circ, 200^\circ]$ or $[360^\circ, 600^\circ]$ for $1 \sigma$ agreement with $\eta_{B_{CMB}}$.

We have further shown successful leptogenesis within the framework of the Neutrino Option scenario is possible when the requisite CP violation in leptogenesis is provided exclusively by the Dirac or Majorana low-energy CP violation phases of the PMNS matrix. This is possible only at the upper bound of the viable mass range and provides a stark contrast with leptogenesis at the lower bounds where the low-energy PMNS phases were largely irrelevant.

\section*{Note}

While this work was in process, we became aware of related work by Vedran Brdar, Alexan-
der J. Helmboldt, Sho Iwamoto and Kai Schmitz, who find analogous results on the viability of leptogenesis in the Neutrino Option~\cite{Brdar:2019iem}. Previous work by some of the same authors, that identified a possible embedding of the Neutrino Option in a conformal theory,should also be acknowledged here \cite{Brdar:2018vjq,Brdar:2018num}. Finally, we would like to thank Vedran Brdar, Alexander J. Helmboldt, Sho Iwamoto and Kai Schmitz for informing us they were undertaking similar work and for being communicative and collegial.

\acknowledgments
We would like to thank Bhupal Dev and Peter Millington for helpful discussions regarding resonant leptogenesis. We thank Vedran Brdar, Alexander J. Helmboldt, Sho Iwamoto and
Kai Schmitz who informed us they were undertaking similar work and were communicative
and collegial.
K.M. and S.P. acknowledge the (partial) support from the European Research Council under the European Union Seventh Framework Programme (FP/2007-2013) / ERC Grant NuMass agreement n. [617143].  S.P. would like to acknowledge partial support from the Wolfson Foundation and the Royal Society, and also thanks SISSA for support and hospitality during part of this work. S.P. and S.T.P. acknowledge partial support from the European Unions Horizon 2020 research and innovation programme under the Marie Sklodowska Curie grant agreements No 690575 (RISE InvisiblesPlus) and No 674896 (ITN ELUSIVE). The work of S.T.P. was supported in part by the INFN program on Theoretical Astroparticle Physics (TASP) and by the World Premier International Research Center Initiative (WPI Initiative), MEXT, Japan. 
 This manuscript has been authored by Fermi Research Alliance, LLC under Contract No. DE-AC02-07CH11359 with the U.S. Department of Energy, Office of Science, Office of High Energy Physics. 

\bibliography{lepbib}{}

\providecommand{\href}[2]{#2}\begingroup\raggedright\begin{thebibliography}{10}

\bibitem{Minkowski:1977sc}
P.~Minkowski, \emph{{$\mu \to e\gamma$ at a Rate of One Out of $10^{9}$ Muon
  Decays?}}, \href{https://doi.org/10.1016/0370-2693(77)90435-X}{\emph{Phys.
  Lett.} {\bfseries B67} (1977) 421--428}.

\bibitem{Yanagida:1979as}
T.~Yanagida, \emph{{HORIZONTAL SYMMETRY AND MASSES OF NEUTRINOS}}, {\emph{Conf.
  Proc.} {\bfseries C7902131} (1979) 95--99}.

\bibitem{GellMann:1980vs}
M.~Gell-Mann, P.~Ramond and R.~Slansky, \emph{{Complex Spinors and Unified
  Theories}}, {\emph{Conf. Proc.} {\bfseries C790927} (1979) 315--321},
  [\href{https://arxiv.org/abs/1306.4669}{{\ttfamily 1306.4669}}].

\bibitem{Mohapatra:1979ia}
R.~N. Mohapatra and G.~Senjanovic, \emph{{Neutrino Mass and Spontaneous Parity
  Violation}}, \href{https://doi.org/10.1103/PhysRevLett.44.912}{\emph{Phys.
  Rev. Lett.} {\bfseries 44} (1980) 912}.

\bibitem{Fukugita:1986hr}
M.~Fukugita and T.~Yanagida, \emph{{Baryogenesis Without Grand Unification}},
  \href{https://doi.org/10.1016/0370-2693(86)91126-3}{\emph{Phys. Lett.}
  {\bfseries B174} (1986) 45--47}.

\bibitem{Buchmuller:1996pa}
W.~Buchmuller and M.~Plumacher, \emph{{Baryon asymmetry and neutrino mixing}},
  \href{https://doi.org/10.1016/S0370-2693(96)01232-4}{\emph{Phys. Lett.}
  {\bfseries B389} (1996) 73--77},
  [\href{https://arxiv.org/abs/hep-ph/9608308}{{\ttfamily hep-ph/9608308}}].

\bibitem{Buchmuller:2004nz}
W.~Buchmuller, P.~Di~Bari and M.~Plumacher, \emph{{Leptogenesis for
  pedestrians}}, \href{https://doi.org/10.1016/j.aop.2004.02.003}{\emph{Annals
  Phys.} {\bfseries 315} (2005) 305--351},
  [\href{https://arxiv.org/abs/hep-ph/0401240}{{\ttfamily hep-ph/0401240}}].

\bibitem{Buchmuller:2005eh}
W.~Buchmuller, R.~D. Peccei and T.~Yanagida, \emph{{Leptogenesis as the origin
  of matter}},
  \href{https://doi.org/10.1146/annurev.nucl.55.090704.151558}{\emph{Ann. Rev.
  Nucl. Part. Sci.} {\bfseries 55} (2005) 311--355},
  [\href{https://arxiv.org/abs/hep-ph/0502169}{{\ttfamily hep-ph/0502169}}].

\bibitem{Pilaftsis:1997jf}
A.~Pilaftsis, \emph{{CP violation and baryogenesis due to heavy Majorana
  neutrinos}}, \href{https://doi.org/10.1103/PhysRevD.56.5431}{\emph{Phys.
  Rev.} {\bfseries D56} (1997) 5431--5451},
  [\href{https://arxiv.org/abs/hep-ph/9707235}{{\ttfamily hep-ph/9707235}}].

\bibitem{Sakharov:1967dj}
A.~D. Sakharov, \emph{{Violation of CP Invariance, c Asymmetry, and Baryon
  Asymmetry of the Universe}},
  \href{https://doi.org/10.1070/PU1991v034n05ABEH002497}{\emph{Pisma Zh. Eksp.
  Teor. Fiz.} {\bfseries 5} (1967) 32--35}.

\bibitem{Vissani:1997ys}
F.~Vissani, \emph{{Do experiments suggest a hierarchy problem?}},
  \href{https://doi.org/10.1103/PhysRevD.57.7027}{\emph{Phys. Rev.} {\bfseries
  D57} (1998) 7027--7030},
  [\href{https://arxiv.org/abs/hep-ph/9709409}{{\ttfamily hep-ph/9709409}}].

\bibitem{Clarke:2015gwa}
J.~D. Clarke, R.~Foot and R.~R. Volkas, \emph{{Electroweak naturalness in the
  three-flavor type I seesaw model and implications for leptogenesis}},
  \href{https://doi.org/10.1103/PhysRevD.91.073009}{\emph{Phys. Rev.}
  {\bfseries D91} (2015) 073009},
  [\href{https://arxiv.org/abs/1502.01352}{{\ttfamily 1502.01352}}].

\bibitem{Davidson:2002qv}
S.~Davidson and A.~Ibarra, \emph{{A Lower bound on the right-handed neutrino
  mass from leptogenesis}},
  \href{https://doi.org/10.1016/S0370-2693(02)01735-5}{\emph{Phys. Lett.}
  {\bfseries B535} (2002) 25--32},
  [\href{https://arxiv.org/abs/hep-ph/0202239}{{\ttfamily hep-ph/0202239}}].

\bibitem{Buchmuller:2002rq}
W.~Buchmuller, P.~Di~Bari and M.~Plumacher, \emph{{Cosmic microwave background,
  matter - antimatter asymmetry and neutrino masses}},
  \href{https://doi.org/10.1016/S0550-3213(02)00737-X,
  10.1016/j.nuclphysb.2007.11.030}{\emph{Nucl. Phys.} {\bfseries B643} (2002)
  367--390}, [\href{https://arxiv.org/abs/hep-ph/0205349}{{\ttfamily
  hep-ph/0205349}}].

\bibitem{Ellis:2002xg}
J.~R. Ellis and M.~Raidal, \emph{{Leptogenesis and the violation of lepton
  number and CP at low-energies}},
  \href{https://doi.org/10.1016/S0550-3213(02)00753-8}{\emph{Nucl. Phys.}
  {\bfseries B643} (2002) 229--246},
  [\href{https://arxiv.org/abs/hep-ph/0206174}{{\ttfamily hep-ph/0206174}}].

\bibitem{Brivio:2017dfq}
I.~Brivio and M.~Trott, \emph{{Radiatively Generating the Higgs Potential and
  Electroweak Scale via the Seesaw Mechanism}},
  \href{https://doi.org/10.1103/PhysRevLett.119.141801}{\emph{Phys. Rev. Lett.}
  {\bfseries 119} (2017) 141801},
  [\href{https://arxiv.org/abs/1703.10924}{{\ttfamily 1703.10924}}].

\bibitem{Brivio:2018rzm}
I.~Brivio and M.~Trott, \emph{{Examining the neutrino option}},
  \href{https://doi.org/10.1007/JHEP02(2019)107}{\emph{JHEP} {\bfseries 02}
  (2019) 107}, [\href{https://arxiv.org/abs/1809.03450}{{\ttfamily
  1809.03450}}].

\bibitem{Wolfenstein:1981kw}
L.~Wolfenstein, \emph{{Different Varieties of Massive Dirac Neutrinos}},
  \href{https://doi.org/10.1016/0550-3213(81)90096-1}{\emph{Nucl. Phys.}
  {\bfseries B186} (1981) 147--152}.

\bibitem{Petcov:1982ya}
S.~T. Petcov, \emph{{On Pseudodirac Neutrinos, Neutrino Oscillations and
  Neutrinoless Double beta Decay}},
  \href{https://doi.org/10.1016/0370-2693(82)91246-1}{\emph{Phys. Lett.}
  {\bfseries 110B} (1982) 245--249}.

\bibitem{PDG2018}
K.~Nakamura and S.~Petcov, \emph{{(Particle Data Group)}}, {\emph{Phys. Rev.}
  {\bfseries D98} (2018) 030001}.

\bibitem{Bilenky:1980cx}
S.~M. Bilenky, J.~Hosek and S.~T. Petcov, \emph{{On Oscillations of Neutrinos
  with Dirac and Majorana Masses}},
  \href{https://doi.org/10.1016/0370-2693(80)90927-2}{\emph{Phys. Lett.}
  {\bfseries 94B} (1980) 495--498}.

\bibitem{Esteban:2018azc}
I.~Esteban, M.~C. Gonzalez-Garcia, A.~Hernandez-Cabezudo, M.~Maltoni and
  T.~Schwetz, \emph{{Global analysis of three-flavour neutrino oscillations:
  synergies and tensions in the determination of $\theta_{23}, \delta_{CP}$,
  and the mass ordering}},
  \href{https://doi.org/10.1007/JHEP01(2019)106}{\emph{JHEP} {\bfseries 01}
  (2019) 106}, [\href{https://arxiv.org/abs/1811.05487}{{\ttfamily
  1811.05487}}].

\bibitem{Casas:2001sr}
J.~A. Casas and A.~Ibarra, \emph{{Oscillating neutrinos and muon ---> e,
  gamma}}, \href{https://doi.org/10.1016/S0550-3213(01)00475-8}{\emph{Nucl.
  Phys.} {\bfseries B618} (2001) 171--204},
  [\href{https://arxiv.org/abs/hep-ph/0103065}{{\ttfamily hep-ph/0103065}}].

\bibitem{Aghanim:2018eyx}
{\scshape Planck} collaboration, N.~Aghanim et~al., \emph{{Planck 2018 results.
  VI. Cosmological parameters}},
  \href{https://arxiv.org/abs/1807.06209}{{\ttfamily 1807.06209}}.

\bibitem{Molinaro:2008rg}
E.~Molinaro and S.~T. Petcov, \emph{{The Interplay Between the 'Low' and 'High'
  Energy CP-Violation in Leptogenesis}},
  \href{https://doi.org/10.1140/epjc/s10052-009-0985-3}{\emph{Eur. Phys. J.}
  {\bfseries C61} (2009) 93--109},
  [\href{https://arxiv.org/abs/0803.4120}{{\ttfamily 0803.4120}}].

\bibitem{Brdar:2018vjq}
V.~Brdar, Y.~Emonds, A.~J. Helmboldt and M.~Lindner, \emph{{Conformal
  Realization of the Neutrino Option}},
  \href{https://doi.org/10.1103/PhysRevD.99.055014}{\emph{Phys. Rev.}
  {\bfseries D99} (2019) 055014},
  [\href{https://arxiv.org/abs/1807.11490}{{\ttfamily 1807.11490}}].

\bibitem{Brdar:2018num}
V.~Brdar, A.~J. Helmboldt and J.~Kubo, \emph{{Gravitational Waves from
  First-Order Phase Transitions: LIGO as a Window to Unexplored Seesaw
  Scales}}, \href{https://doi.org/10.1088/1475-7516/2019/02/021}{\emph{JCAP}
  {\bfseries 1902} (2019) 021},
  [\href{https://arxiv.org/abs/1810.12306}{{\ttfamily 1810.12306}}].

\bibitem{Buttazzo:2013uya}
D.~Buttazzo, G.~Degrassi, P.~P. Giardino, G.~F. Giudice, F.~Sala, A.~Salvio
  et~al., \emph{{Investigating the near-criticality of the Higgs boson}},
  \href{https://doi.org/10.1007/JHEP12(2013)089}{\emph{JHEP} {\bfseries 12}
  (2013) 089}, [\href{https://arxiv.org/abs/1307.3536}{{\ttfamily 1307.3536}}].

\bibitem{Patrignani:2016xqp}
{\scshape Particle Data Group} collaboration, C.~Patrignani et~al.,
  \emph{{Review of Particle Physics}},
  \href{https://doi.org/10.1088/1674-1137/40/10/100001}{\emph{Chin. Phys.}
  {\bfseries C40} (2016) 100001}.

\bibitem{Ade:2015xua}
{\scshape Planck} collaboration, P.~A.~R. Ade et~al., \emph{{Planck 2015
  results. XIII. Cosmological parameters}},
  \href{https://doi.org/10.1051/0004-6361/201525830}{\emph{Astron. Astrophys.}
  {\bfseries 594} (2016) A13},
  [\href{https://arxiv.org/abs/1502.01589}{{\ttfamily 1502.01589}}].

\bibitem{Blanchet:2011xq}
S.~Blanchet, P.~Di~Bari, D.~A. Jones and L.~Marzola, \emph{{Leptogenesis with
  heavy neutrino flavours: from density matrix to Boltzmann equations}},
  \href{https://doi.org/10.1088/1475-7516/2013/01/041}{\emph{JCAP} {\bfseries
  1301} (2013) 041}, [\href{https://arxiv.org/abs/1112.4528}{{\ttfamily
  1112.4528}}].

\bibitem{Pilaftsis:2003gt}
A.~Pilaftsis and T.~E.~J. Underwood, \emph{{Resonant leptogenesis}},
  \href{https://doi.org/10.1016/j.nuclphysb.2004.05.029}{\emph{Nucl. Phys.}
  {\bfseries B692} (2004) 303--345},
  [\href{https://arxiv.org/abs/hep-ph/0309342}{{\ttfamily hep-ph/0309342}}].

\bibitem{Dev:2014laa}
P.~S. Bhupal~Dev, P.~Millington, A.~Pilaftsis and D.~Teresi, \emph{{Flavour
  Covariant Transport Equations: an Application to Resonant Leptogenesis}},
  \href{https://doi.org/10.1016/j.nuclphysb.2014.06.020}{\emph{Nucl. Phys.}
  {\bfseries B886} (2014) 569--664},
  [\href{https://arxiv.org/abs/1404.1003}{{\ttfamily 1404.1003}}].

\bibitem{Dev:2014wsa}
P.~S. Bhupal~Dev, P.~Millington, A.~Pilaftsis and D.~Teresi,
  \emph{{KadanoffÐBaym approach to flavour mixing and oscillations in resonant
  leptogenesis}},
  \href{https://doi.org/10.1016/j.nuclphysb.2014.12.003}{\emph{Nucl. Phys.}
  {\bfseries B891} (2015) 128--158},
  [\href{https://arxiv.org/abs/1410.6434}{{\ttfamily 1410.6434}}].

\bibitem{Bambhaniya:2016rbb}
G.~Bambhaniya, P.~S. Bhupal~Dev, S.~Goswami, S.~Khan and W.~Rodejohann,
  \emph{{Naturalness, Vacuum Stability and Leptogenesis in the Minimal Seesaw
  Model}}, \href{https://doi.org/10.1103/PhysRevD.95.095016}{\emph{Phys. Rev.}
  {\bfseries D95} (2017) 095016},
  [\href{https://arxiv.org/abs/1611.03827}{{\ttfamily 1611.03827}}].

\bibitem{Moffat:2018wke}
K.~Moffat, S.~Pascoli, S.~T. Petcov, H.~Schulz and J.~Turner,
  \emph{{Three-Flavoured Non-Resonant Leptogenesis at Intermediate Scales}},
  \href{https://arxiv.org/abs/1804.05066}{{\ttfamily 1804.05066}}.

\bibitem{Blanchet:2008pw}
S.~Blanchet and P.~Di~Bari, \emph{{New aspects of leptogenesis bounds}},
  \href{https://doi.org/10.1016/j.nuclphysb.2008.08.026}{\emph{Nucl. Phys.}
  {\bfseries B807} (2009) 155--187},
  [\href{https://arxiv.org/abs/0807.0743}{{\ttfamily 0807.0743}}].

\bibitem{Antusch:2009gn}
S.~Antusch, S.~Blanchet, M.~Blennow and E.~Fernandez-Martinez,
  \emph{{Non-unitary Leptonic Mixing and Leptogenesis}},
  \href{https://doi.org/10.1007/JHEP01(2010)017}{\emph{JHEP} {\bfseries 01}
  (2010) 017}, [\href{https://arxiv.org/abs/0910.5957}{{\ttfamily 0910.5957}}].

\bibitem{Pilaftsis:2005rv}
A.~Pilaftsis and T.~E.~J. Underwood, \emph{{Electroweak-scale resonant
  leptogenesis}}, \href{https://doi.org/10.1103/PhysRevD.72.113001}{\emph{Phys.
  Rev.} {\bfseries D72} (2005) 113001},
  [\href{https://arxiv.org/abs/hep-ph/0506107}{{\ttfamily hep-ph/0506107}}].

\bibitem{Hambye:2001eu}
T.~Hambye, \emph{{Leptogenesis at the TeV scale}},
  \href{https://doi.org/10.1016/S0550-3213(02)00293-6}{\emph{Nucl. Phys.}
  {\bfseries B633} (2002) 171--192},
  [\href{https://arxiv.org/abs/hep-ph/0111089}{{\ttfamily hep-ph/0111089}}].

\bibitem{Hambye:2004jf}
T.~Hambye, J.~March-Russell and S.~M. West, \emph{{TeV scale resonant
  leptogenesis from supersymmetry breaking}},
  \href{https://doi.org/10.1088/1126-6708/2004/07/070}{\emph{JHEP} {\bfseries
  07} (2004) 070}, [\href{https://arxiv.org/abs/hep-ph/0403183}{{\ttfamily
  hep-ph/0403183}}].

\bibitem{Cirigliano:2006nu}
V.~Cirigliano, G.~Isidori and V.~Porretti, \emph{{CP violation and Leptogenesis
  in models with Minimal Lepton Flavour Violation}},
  \href{https://doi.org/10.1016/j.nuclphysb.2006.11.015}{\emph{Nucl. Phys.}
  {\bfseries B763} (2007) 228--246},
  [\href{https://arxiv.org/abs/hep-ph/0607068}{{\ttfamily hep-ph/0607068}}].

\bibitem{Xing:2006ms}
Z.-z. Xing and S.~Zhou, \emph{{Tri-bimaximal Neutrino Mixing and
  Flavor-dependent Resonant Leptogenesis}},
  \href{https://doi.org/10.1016/j.physletb.2007.08.009}{\emph{Phys. Lett.}
  {\bfseries B653} (2007) 278--287},
  [\href{https://arxiv.org/abs/hep-ph/0607302}{{\ttfamily hep-ph/0607302}}].

\bibitem{Branco:2006hz}
G.~C. Branco, A.~J. Buras, S.~Jager, S.~Uhlig and A.~Weiler, \emph{{Another
  look at minimal lepton flavour violation, $l_i \to l_{j\gamma}$,
  leptogenesis, and the ratio $M_\nu / \Lambda_{LFV}$}},
  \href{https://doi.org/10.1088/1126-6708/2007/09/004}{\emph{JHEP} {\bfseries
  09} (2007) 004}, [\href{https://arxiv.org/abs/hep-ph/0609067}{{\ttfamily
  hep-ph/0609067}}].

\bibitem{Chun:2007vh}
E.~J. Chun and K.~Turzynski, \emph{{Quasi-degenerate neutrinos and leptogenesis
  from L(mu) - L(tau)}},
  \href{https://doi.org/10.1103/PhysRevD.76.053008}{\emph{Phys. Rev.}
  {\bfseries D76} (2007) 053008},
  [\href{https://arxiv.org/abs/hep-ph/0703070}{{\ttfamily hep-ph/0703070}}].

\bibitem{Kitabayashi:2007bs}
T.~Kitabayashi, \emph{{Remark on the minimal seesaw model and leptogenesis with
  tri/bi-maximal mixing}},
  \href{https://doi.org/10.1103/PhysRevD.76.033002}{\emph{Phys. Rev.}
  {\bfseries D76} (2007) 033002},
  [\href{https://arxiv.org/abs/hep-ph/0703303}{{\ttfamily hep-ph/0703303}}].

\bibitem{Lopez-Pavon:2015cga}
J.~Lopez-Pavon, E.~Molinaro and S.~T. Petcov, \emph{{Radiative Corrections to
  Light Neutrino Masses in Low Scale Type I Seesaw Scenarios and Neutrinoless
  Double Beta Decay}},
  \href{https://doi.org/10.1007/JHEP11(2015)030}{\emph{JHEP} {\bfseries 11}
  (2015) 030}, [\href{https://arxiv.org/abs/1506.05296}{{\ttfamily
  1506.05296}}].

\bibitem{Casas:1999tg}
J.~A. Casas, J.~R. Espinosa, A.~Ibarra and I.~Navarro, \emph{{General RG
  equations for physical neutrino parameters and their phenomenological
  implications}},
  \href{https://doi.org/10.1016/S0550-3213(99)00781-6}{\emph{Nucl. Phys.}
  {\bfseries B573} (2000) 652--684},
  [\href{https://arxiv.org/abs/hep-ph/9910420}{{\ttfamily hep-ph/9910420}}].

\bibitem{Antusch:2003kp}
S.~Antusch, J.~Kersten, M.~Lindner and M.~Ratz, \emph{{Running neutrino masses,
  mixings and CP phases: Analytical results and phenomenological
  consequences}},
  \href{https://doi.org/10.1016/j.nuclphysb.2003.09.050}{\emph{Nucl. Phys.}
  {\bfseries B674} (2003) 401--433},
  [\href{https://arxiv.org/abs/hep-ph/0305273}{{\ttfamily hep-ph/0305273}}].

\bibitem{Feroz:2007kg}
F.~Feroz and M.~P. Hobson, \emph{{Multimodal nested sampling: an efficient and
  robust alternative to MCMC methods for astronomical data analysis}},
  \href{https://doi.org/10.1111/j.1365-2966.2007.12353.x}{\emph{Mon. Not. Roy.
  Astron. Soc.} {\bfseries 384} (2008) 449},
  [\href{https://arxiv.org/abs/0704.3704}{{\ttfamily 0704.3704}}].

\bibitem{Feroz:2008xx}
F.~Feroz, M.~P. Hobson and M.~Bridges, \emph{{MultiNest: an efficient and
  robust Bayesian inference tool for cosmology and particle physics}},
  \href{https://doi.org/10.1111/j.1365-2966.2009.14548.x}{\emph{Mon. Not. Roy.
  Astron. Soc.} {\bfseries 398} (2009) 1601--1614},
  [\href{https://arxiv.org/abs/0809.3437}{{\ttfamily 0809.3437}}].

\bibitem{Feroz:2013hea}
F.~Feroz, M.~P. Hobson, E.~Cameron and A.~N. Pettitt, \emph{{Importance Nested
  Sampling and the MultiNest Algorithm}},
  \href{https://arxiv.org/abs/1306.2144}{{\ttfamily 1306.2144}}.

\bibitem{Mohapatra:1986aw}
R.~N. Mohapatra, \emph{{Mechanism for Understanding Small Neutrino Mass in
  Superstring Theories}},
  \href{https://doi.org/10.1103/PhysRevLett.56.561}{\emph{Phys. Rev. Lett.}
  {\bfseries 56} (1986) 561--563}.

\bibitem{Mohapatra:1986bd}
R.~N. Mohapatra and J.~W.~F. Valle, \emph{{Neutrino Mass and Baryon Number
  Nonconservation in Superstring Models}},
  \href{https://doi.org/10.1103/PhysRevD.34.1642}{\emph{Phys. Rev.} {\bfseries
  D34} (1986) 1642}.

\bibitem{Bernabeu:1987gr}
J.~{Bernab\'eu}, A.~Santamaria, J.~Vidal, A.~Mendez and J.~W.~F. Valle,
  \emph{{Lepton Flavor Nonconservation at High-Energies in a Superstring
  Inspired Standard Model}},
  \href{https://doi.org/10.1016/0370-2693(87)91100-2}{\emph{Phys. Lett.}
  {\bfseries B187} (1987) 303}.

\bibitem{Pilaftsis:1991ug}
A.~Pilaftsis, \emph{{Radiatively induced neutrino masses and large Higgs
  neutrino couplings in the standard model with Majorana fields}},
  \href{https://doi.org/10.1007/BF01482590}{\emph{Z. Phys.} {\bfseries C55}
  (1992) 275--282}, [\href{https://arxiv.org/abs/hep-ph/9901206}{{\ttfamily
  hep-ph/9901206}}].

\bibitem{Ilakovac:1994kj}
A.~Ilakovac and A.~Pilaftsis, \emph{{Flavor violating charged lepton decays in
  seesaw-type models}},
  \href{https://doi.org/10.1016/0550-3213(94)00567-X}{\emph{Nucl. Phys.}
  {\bfseries B437} (1995) 491},
  [\href{https://arxiv.org/abs/hep-ph/9403398}{{\ttfamily hep-ph/9403398}}].

\bibitem{Akhmedov:1995ip}
E.~K. Akhmedov, M.~Lindner, E.~Schnapka and J.~W.~F. Valle, \emph{{Left-right
  symmetry breaking in NJL approach}},
  \href{https://doi.org/10.1016/0370-2693(95)01504-3}{\emph{Phys. Lett.}
  {\bfseries B368} (1996) 270--280},
  [\href{https://arxiv.org/abs/hep-ph/9507275}{{\ttfamily hep-ph/9507275}}].

\bibitem{Akhmedov:1995vm}
E.~K. Akhmedov, M.~Lindner, E.~Schnapka and J.~W.~F. Valle, \emph{{Dynamical
  left-right symmetry breaking}},
  \href{https://doi.org/10.1103/PhysRevD.53.2752}{\emph{Phys. Rev.} {\bfseries
  D53} (1996) 2752--2780},
  [\href{https://arxiv.org/abs/hep-ph/9509255}{{\ttfamily hep-ph/9509255}}].

\bibitem{Abada:2010ym}
A.~Abada, G.~Bhattacharyya, D.~Das and C.~Weiland, \emph{{A possible connection
  between neutrino mass generation and the lightness of a NMSSM pseudoscalar}},
  \href{https://doi.org/10.1016/j.physletb.2011.05.020}{\emph{Phys. Lett.}
  {\bfseries B700} (2011) 351--355},
  [\href{https://arxiv.org/abs/1011.5037}{{\ttfamily 1011.5037}}].

\bibitem{Abada:2011hm}
A.~Abada, D.~Das and C.~Weiland, \emph{{Enhanced Higgs Mediated Lepton Flavour
  Violating Processes in the Supersymmetric Inverse Seesaw Model}},
  \href{https://doi.org/10.1007/JHEP03(2012)100}{\emph{JHEP} {\bfseries 03}
  (2012) 100}, [\href{https://arxiv.org/abs/1111.5836}{{\ttfamily 1111.5836}}].

\bibitem{Alonso:2016onw}
R.~Alonso, E.~Fernandez~Martinez, M.~B. Gavela, B.~Grinstein, L.~Merlo and
  P.~Quilez, \emph{{Gauged Lepton Flavour}},
  \href{https://doi.org/10.1007/JHEP12(2016)119}{\emph{JHEP} {\bfseries 12}
  (2016) 119}, [\href{https://arxiv.org/abs/1609.05902}{{\ttfamily
  1609.05902}}].

\bibitem{Gavela:2009cd}
M.~B. Gavela, T.~Hambye, D.~Hernandez and P.~Hernandez, \emph{{Minimal Flavour
  Seesaw Models}},
  \href{https://doi.org/10.1088/1126-6708/2009/09/038}{\emph{JHEP} {\bfseries
  09} (2009) 038}, [\href{https://arxiv.org/abs/0906.1461}{{\ttfamily
  0906.1461}}].

\bibitem{Dias:2011sq}
A.~G. Dias, C.~A. de~S.~Pires and P.~S.~R. da~Silva, \emph{{How the Inverse
  See-Saw Mechanism Can Reveal Itself Natural, Canonical and Independent of the
  Right-Handed Neutrino Mass}},
  \href{https://doi.org/10.1103/PhysRevD.84.053011}{\emph{Phys. Rev.}
  {\bfseries D84} (2011) 053011},
  [\href{https://arxiv.org/abs/1107.0739}{{\ttfamily 1107.0739}}].

\bibitem{Bazzocchi:2010dt}
F.~Bazzocchi, \emph{{Minimal Dynamical Inverse See Saw}},
  \href{https://doi.org/10.1103/PhysRevD.83.093009}{\emph{Phys. Rev.}
  {\bfseries D83} (2011) 093009},
  [\href{https://arxiv.org/abs/1011.6299}{{\ttfamily 1011.6299}}].

\bibitem{Ma:2009gu}
E.~Ma, \emph{{Radiative inverse seesaw mechanism for nonzero neutrino mass}},
  \href{https://doi.org/10.1103/PhysRevD.80.013013}{\emph{Phys. Rev.}
  {\bfseries D80} (2009) 013013},
  [\href{https://arxiv.org/abs/0904.4450}{{\ttfamily 0904.4450}}].

\bibitem{Moffat:2018smo}
K.~Moffat, S.~Pascoli, S.~T. Petcov and J.~Turner, \emph{{Leptogenesis from Low
  Energy $CP$ Violation}},
  \href{https://doi.org/10.1007/JHEP03(2019)034}{\emph{JHEP} {\bfseries 03}
  (2019) 034}, [\href{https://arxiv.org/abs/1809.08251}{{\ttfamily
  1809.08251}}].

\bibitem{Pascoli:2006ie}
S.~Pascoli, S.~T. Petcov and A.~Riotto, \emph{{Connecting low energy leptonic
  CP-violation to leptogenesis}},
  \href{https://doi.org/10.1103/PhysRevD.75.083511}{\emph{Phys. Rev.}
  {\bfseries D75} (2007) 083511},
  [\href{https://arxiv.org/abs/hep-ph/0609125}{{\ttfamily hep-ph/0609125}}].

\bibitem{Pascoli:2006ci}
S.~Pascoli, S.~T. Petcov and A.~Riotto, \emph{{Leptogenesis and Low Energy CP
  Violation in Neutrino Physics}},
  \href{https://doi.org/10.1016/j.nuclphysb.2007.02.019}{\emph{Nucl. Phys.}
  {\bfseries B774} (2007) 1--52},
  [\href{https://arxiv.org/abs/hep-ph/0611338}{{\ttfamily hep-ph/0611338}}].

\bibitem{Blanchet:2006be}
S.~Blanchet and P.~Di~Bari, \emph{{Flavor effects on leptogenesis
  predictions}},
  \href{https://doi.org/10.1088/1475-7516/2007/03/018}{\emph{JCAP} {\bfseries
  0703} (2007) 018}, [\href{https://arxiv.org/abs/hep-ph/0607330}{{\ttfamily
  hep-ph/0607330}}].

\bibitem{Branco:2006ce}
G.~C. Branco, R.~Gonzalez~Felipe and F.~R. Joaquim, \emph{{A New bridge between
  leptonic CP violation and leptogenesis}},
  \href{https://doi.org/10.1016/j.physletb.2006.12.060}{\emph{Phys. Lett.}
  {\bfseries B645} (2007) 432--436},
  [\href{https://arxiv.org/abs/hep-ph/0609297}{{\ttfamily hep-ph/0609297}}].

\bibitem{Anisimov:2007mw}
A.~Anisimov, S.~Blanchet and P.~Di~Bari, \emph{{Viability of Dirac phase
  leptogenesis}},
  \href{https://doi.org/10.1088/1475-7516/2008/04/033}{\emph{JCAP} {\bfseries
  0804} (2008) 033}, [\href{https://arxiv.org/abs/0707.3024}{{\ttfamily
  0707.3024}}].

\bibitem{Molinaro:2008cw}
E.~Molinaro and S.~T. Petcov, \emph{{A Case of Subdominant/Suppressed 'High
  Energy' Contribution to the Baryon Asymmetry of the Universe in Flavoured
  Leptogenesis}},
  \href{https://doi.org/10.1016/j.physletb.2008.11.047}{\emph{Phys. Lett.}
  {\bfseries B671} (2009) 60--65},
  [\href{https://arxiv.org/abs/0808.3534}{{\ttfamily 0808.3534}}].

\bibitem{Dolan:2018qpy}
M.~J. Dolan, T.~P. Dutka and R.~R. Volkas, \emph{{Dirac-Phase Thermal
  Leptogenesis in the extended Type-I Seesaw Model}},
  \href{https://doi.org/10.1088/1475-7516/2018/06/012}{\emph{JCAP} {\bfseries
  1806} (2018) 012}, [\href{https://arxiv.org/abs/1802.08373}{{\ttfamily
  1802.08373}}].

\bibitem{Hagedorn:2017wjy}
C.~Hagedorn, R.~N. Mohapatra, E.~Molinaro, C.~C. Nishi and S.~T. Petcov,
  \emph{{CP Violation in the Lepton Sector and Implications for Leptogenesis}},
   \href{https://arxiv.org/abs/1711.02866}{{\ttfamily 1711.02866}}.

\bibitem{Bilenky:1987ty}
S.~M. Bilenky and S.~T. Petcov, \emph{{Massive Neutrinos and Neutrino
  Oscillations}}, \href{https://doi.org/10.1103/RevModPhys.59.671}{\emph{Rev.
  Mod. Phys.} {\bfseries 59} (1987) 671}.

\bibitem{Pascoli:2005zb}
S.~Pascoli, S.~T. Petcov and T.~Schwetz, \emph{{The Absolute neutrino mass
  scale, neutrino mass spectrum, majorana CP-violation and neutrinoless
  double-beta decay}},
  \href{https://doi.org/10.1016/j.nuclphysb.2005.11.003}{\emph{Nucl. Phys.}
  {\bfseries B734} (2006) 24--49},
  [\href{https://arxiv.org/abs/hep-ph/0505226}{{\ttfamily hep-ph/0505226}}].

\bibitem{Brdar:2019iem}
V.~Brdar, A.~J. Helmboldt, S.~Iwamoto and K.~Schmitz, \emph{{Type-I Seesaw as
  the Common Origin of Neutrino Mass, Baryon Asymmetry, and the Electroweak
  Scale}}, \href{https://doi.org/10.1103/PhysRevD.100.075029}{\emph{Phys. Rev.}
  {\bfseries D100} (2019) 075029},
  [\href{https://arxiv.org/abs/1905.12634}{{\ttfamily 1905.12634}}].

\end{thebibliography}\endgroup
\bibliographystyle{JHEP}

\end{document}